\documentclass{aa}  

\usepackage{graphicx}
\usepackage{natbib}
\usepackage[dvipsnames]{xcolor}
\usepackage{amsmath}
\usepackage[switch]{lineno}
\usepackage[normalem]{ulem}
\usepackage{float}
\usepackage{placeins}
\usepackage{dblfloatfix}
\graphicspath{{Figures/}}

\bibpunct{(}{)}{;}{a}{}{,}

\begin{document} 

\title{The influence of bulk composition on the long-term interior-atmosphere evolution of terrestrial exoplanets}
\author{Rob J. Spaargaren \inst{\ref{inst1}}
\and Maxim D. Ballmer \inst{\ref{inst2}}\inst{\ref{inst1}}
\and Dan J. Bower \inst{\ref{inst3}}
\and Caroline Dorn \inst{\ref{inst4}}
\and Paul J. Tackley \inst{\ref{inst1}}}

\institute{Institute of Geophysics, ETH Zurich, Sonneggstrasse 5, 8092 Zurich, Switzerland \label{inst1}
\and Department of Earth Sciences, University College London, Gower Place, London WC1E 6BT, United Kingdom \label{inst2} \and Center for Space and Habitability, University of Bern, Gesellschaftsstrasse 6, 3012 Bern, Switzerland \label{inst3} 
\and Institute for Computational Science, University of Zurich, Winterthurerstrasse 190, 8057 Zurich, Switzerland \label{inst4} } 

\date{Received 03 February 2020 / Accepted 15 September 2020}

\abstract{}{The secondary atmospheres of terrestrial planets form and evolve as a consequence of interaction with the interior over geological time. We aim to quantify the influence of planetary bulk composition on the interior--atmosphere evolution for Earth-sized terrestrial planets to aid in the interpretation of future observations of terrestrial exoplanet atmospheres.}
{We used a geochemical model to determine the major-element composition of planetary interiors (MgO, FeO, and SiO$_2$) following the crystallization of a magma ocean after planet formation, predicting a compositional profile of the interior as an initial condition for our long-term thermal evolution model. Our 1D evolution model predicts the pressure-temperature structure of the interior, which we used to evaluate near-surface melt production and subsequent volatile outgassing. Volatiles are exchanged between the interior and atmosphere according to mass conservation.}
{Based on stellar compositions reported in the Hypatia catalog, we predict that about half of rocky exoplanets have a mantle that convects as a single layer (whole-mantle convection), and the other half exhibit double-layered convection due to the presence of a mid-mantle compositional boundary. Double-layered convection is more likely for planets with high bulk planetary Fe-content and low Mg/Si-ratio. We find that planets with low Mg/Si-ratio tend to cool slowly because their mantle viscosity is high. Accordingly, low-Mg/Si planets also tend to lose volatiles swiftly through extensive melting. Moreover, the dynamic regime of the lithosphere (plate tectonics vs.\ stagnant lid) has a first-order influence on the thermal evolution and volatile cycling. These results suggest that the composition of terrestrial exoplanetary atmospheres can provide information on the dynamic regime of the lithosphere and the thermo-chemical evolution of the interior.}{}

\keywords{Planets and satellites: terrestrial planets - Planets and satellites: atmospheres - Planets and satellites: composition - Planets and satellites: interiors}

\titlerunning{Bulk composition and interior--atmosphere evolution of terrestrial exoplanets}

\maketitle

\section{Introduction} \label{sec:intro}
A major goal of exoplanetary science is to gain an understanding of the formation and evolution of terrestrial (i.e., rocky) planets. Such an effort demands an understanding of their interior dynamics. Our knowledge of planetary evolution is, to a large extent, based on Solar System objects since the availability of detailed observational data of exoplanets remains limited. Constraints on the interior properties of terrestrial exoplanets are mostly derived from mass-radius observations \citep[e.g.,][]{Seager2007,Weiss2014,Dorn2015,Unterborn2016,Unterborn2019}. The data derived from these observations do not follow a single trend in terms of mass--radius space \citep{Rogers2010,Dorn2015,Dorn2018_outgassing}, indicating a broad diversity in interior compositions and structure. The observed bulk density variations can be explained as the result of diversity among the volatile budget, mantle composition, interior structure, or melt fraction \citep{Valencia2006,Seager2007,Wagner2011,Dorn2015,Dorn2018_TRAPPIST,Bower2019}. Thus, a substantial ambiguity remains with regard to determining interior properties based on mass-radius measurements alone and, thus, the host star composition can be used to reduce this ambiguity \citep[e.g.,][]{Rogers2010,Dorn2015,Brugger2017,Santos2017,Hinkel2018}, as it constrains the interior structure and mantle composition. However, a fundamental ambiguity remains and detailed  studies will be needed in the future to address this point.

Complementary constraints on interior properties might be provided by the atmospheres of terrestrial exoplanets since these form and evolve as a consequence of interaction with the interior \citep[e.g.,][]{Foley2016}. Future developments in astronomical instrumentation will allow us to place some constraints on the chemistry of the upper atmosphere for some terrestrial exoplanets (e.g., JWST, ARIEL, E-ELT). Such atmospheric characterizations may provide an opportunity for disentangling the observed mass--radius trends of terrestrial exoplanets. Invariably, this requires an improved understanding of the coupling between the atmosphere and interior.

Comprehending atmosphere-interior coupling requires insights into the effects of bulk planet composition \citep{Unterborn2017_Tect,Dorn2018_outgassing}, which can vary greatly between systems. Variations in the interior composition can stem from differences in the formation environment of planetary systems. The composition of accreted material depends on the local temperature, as well as the bulk composition, of the protoplanetary disk. The bulk composition of the disk varies between exoplanetary systems, as reflected by the variation in host stellar compositions \citep{Bond2010,Carter2012,Moriarty2014,Thiabaud2015,Unterborn2017_MgSi}. Stellar compositions are a good indicator for planetary compositions in terms of the major, refractory, rock-forming elements: Mg, Fe, Si, Al, and Ca because the planet, disk, and star form from the same cloud of gas and dust. Stellar chemical abundances vary widely across the Solar neighborhood \citep{Bensby2005,Valenti2005,Asplund2009,Lodders2009,Hypatia}, indicating a broad variety in exoplanetary budgets of rock-forming elements, which are the main constituents of terrestrial planets. While the exact extent of stellar chemical variability in the Solar neighborhood is still under debate \citep{Brewer2016,Bedell2018}, we can appeal to current estimates in the literature \citep{Hinkel2018}, acknowledging that they will continue to be refined with new observations to come.

The evolution of planetary atmospheres and interiors following planet formation and core-mantle differentiation is characterized by two main phases. The first phase consists of the crystallization of a magma ocean and the formation of a primary atmosphere, consisting of gases escaping the crystallizing magma ocean. The primary atmosphere replaces a primordial H/He-atmosphere that may have been captured from the protoplanetary disk \citep{Hayashi1979}. The primordial atmosphere escapes almost entirely for planets up to super-Earth size \citep{Owen2013,Fulton2017}. During this phase, the central iron core forms in a magma ocean, where metallic iron accumulates at the center of the planet. Metallic iron and silicates are immiscible at these extreme conditions, which causes the dense metallic phase to differentiate from the silicate phase and sink to the planetary core. Elements are preferentially partitioned into the metallic phase (siderophile elements, e.g.,\ Fe, Ni, Au, Mn, and Pt) or the silicate phase (lithophile elements, e.g.,\ Si, Mg, Na, Ca, Al, and U). The core size is determined by the abundances of iron and siderophile elements relative to lithophile elements, as well as the oxygen fugacity \citep{Corgne2008,Rubie2015}.

The magma ocean phase also determines the mineralogy and compositional structure of the silicate mantle that resides above the core. Compositional variations between different mantle layers may resist homogenization by convection, as long as a compositional density contrast occurs between the layers that exceeds thermal buoyancy \citep{Kellogg1999,Elkins2008,Tosi2013,Li2014,Plesa2014}. For a density contrast of more than 200 kg/m$^3$, stable stratification may persist for most of the planet's lifetime. Such stratification may promote double-layered convection, with negligible mass transfer between layers, as opposed to the single-layer (whole-mantle) convection that occurs through most of Earth's history to the present day.

The second phase consists of the long-term evolution of both the atmosphere (now called the secondary atmosphere) and interior, after the interior has mostly solidified. Estimates for the timescale for magma-ocean crystallization on Earth range from thousands \citep{Solomatov2000} to millions \citep{Lebrun2013} of years; so, a rocky planet spends most of its life in the second phase, unless it is extremely close to its host star \citep[see e.g.][]{Leger2011}. Thermal evolution of planets with solid interiors is dictated by the viscous creep of solid materials, which manifests as mantle convection. Bulk composition and mantle mineralogy affect the transport and material properties of the mantle, such as density and viscosity \citep{Takeda1998,Yamazaki2001}, and melting behavior \citep{Hirschmann2000,Kiefer2015}. Additionally, the budget of radiogenic elements varies among stars \citep{Frank2014,Unterborn2015,Botelho2019} and determines the heat production in planetary mantles, which also moderates the style of interior convection.

During the second phase, the secondary atmosphere evolves through interaction with the now-solid interior \citep{Sleep2001,Franck2002,Gaillard2014,Foley2018,Laneuville2018} and by loss to space \citep{Zahnle2017}. This interaction involves mantle outgassing (transfer of volatiles, such as H$_2$O and CO$_2$, from rocky interior to atmosphere) and ingassing (transfer of volatiles from atmosphere to rocky interior). The dominant styles of outgassing and ingassing depend on the dynamic regime of the near-surface (i.e., plate tectonics or stagnant lid; see Figs.\ 10 and 11 in \cite{Foley2016}). Ingassing likely occurs at a higher rate for planets in the plate tectonics regime \citep{Tosi2017}, where ingassing occurs by transport of crustal material into the mantle by subduction \citep{Rupke2004}. Outgassing occurs by melt production in the interior and the escape of volatiles by volcanism \citep{Dixon1995,Fuentes2019}. The chemistry of outgassed volatiles depends on the interior state, mainly the oxygen fugacity \citep[e.g.,][]{Holloway1994}, the planetary volatile budget, and atmospheric conditions \citep{Gaillard2014}. These connections between the atmosphere and interior lend support to the idea that an improved understanding of the interior can be gained through observations of planetary atmospheres.

Future observations of terrestrial exoplanet upper atmospheres may give us an unprecedented opportunity to place constraints on the interior composition and dynamics of rocky exoplanets. In order to exploit this opportunity, however, we must continue to advance our comprehension of the interplay between the atmospheric and interior evolution of terrestrial planets. The current generation of coupled atmosphere-interior models have expanded our knowledge of several aspects of atmosphere-interior interaction, such as outgassing from a magma ocean \citep[e.g.,][]{Elkins2008}, volcanic outgassing chemistry \citep[e.g.][]{Gaillard2014}, and the differences between dynamic regimes \citep[e.g.,][]{Tosi2017}. However, these models do not account for the variable bulk composition of terrestrial exoplanets and, thus, they do not have the capacity to assess the influence of bulk planet composition on atmosphere-interior interaction.

\section{Methodology} \label{sec:Metho}
We built a multi-component model that captures the primary aspects of atmosphere--interior coupling for terrestrial planets for a variety of bulk compositions, based on stellar composition data (Sect.~\ref{ssec:Metho_composition}). Our model includes a 1D geochemical model simulating magma ocean crystallization, which determines the initial compositional profile of the interior for each considered bulk composition (Sect.~\ref{ssec:Metho_MO_cryst}). On the basis of this initial condition, we simulate the long-term thermal evolution of the planet using a 1D parametrized convection model (Sect.~\ref{ssec:Metho_Interior}). In the final model component, the efficiency of melt production as a function of thermal evolution is calculated, which is considered as a proxy for outgassing. This model component is a box model that tracks the exchange of volatiles between the interior and atmosphere (Sect.~\ref{ssec:Metho_Volatiles}). We consider water as the representative volatile and we  limit our consideration to planets with the same mass, radius, and core size as Earth.

\subsection{Bulk composition} \label{ssec:Metho_composition}
We define bulk composition in terms of the major rock-forming compounds: FeO, MgO, and SiO$_2$. These are the most abundant compounds in the silicate Earth \citep{Javoy2010} and they are the most abundant refractory elements in most stars in the Solar neighborhood \citep{Hypatia}. Here, we assume that molar ratios of Mg/Si and Fe/Mg are comparable to the host star, in accordance with planetary formation models \citep[e.g.,][]{Bond2010,Carter2012,Thiabaud2015}. We consider only Mg, Fe, and Si since their condensation temperatures and fractionation behavior between the protoplanetary disk and planets are comparable and the corresponding mineral compounds (MgO, FeO, SiO$_2$) have the most significant compositional effect on mantle rheological properties \citep[e.g.,][]{Yamazaki2001}. Moreover, the abundances and properties of mineral phases that are stabilized by other, less abundant compounds are not well-constrained at lower mantle pressures.

\begin{figure}
   \centering
   \includegraphics[width=\hsize]{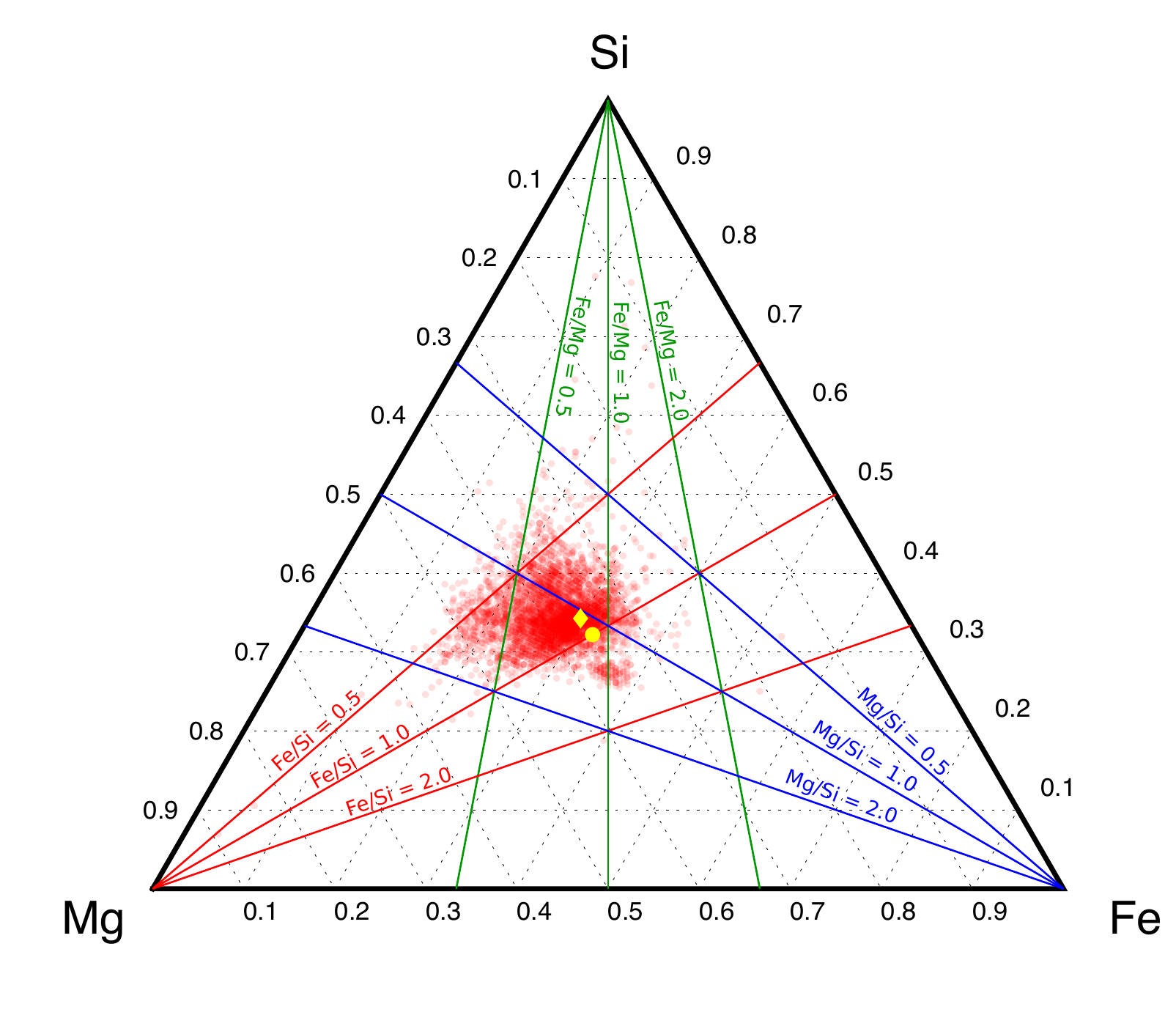} 
      \caption{Compositional diversity of stars in the Solar neighborhood in the molar Mg-Fe-Si system, as recorded in the Hypatia Catalog \citep{Hypatia}. Each red dot represents a star, while the yellow triangle represents the Solar composition from \cite{Asplund2005}, which was also used for converting abundances from the Hypatia catalog. The yellow dot shows Earth's composition \citep{McDonough2003}.}
      \label{fig:Hypatia_comp}
\end{figure}

We used a catalog of stellar compositions in the Solar neighborhood, namely, the Hypatia catalog \citep{Hypatia}, as a proxy for terrestrial exoplanet composition in terms of the molar ratios Mg/Si and Fe/Mg. This catalog maps out compositions of FGK-type stars within 150 pc of the Sun. The majority of stellar compositions display molar Mg/Si-ratios between 0.7 and 1.5, and molar Fe/Mg-ratios between 0.5 and 1.0 (Fig.\ \ref{fig:Hypatia_comp}). Within this range, the Solar composition is average in terms of Mg/Si, but near the high end of Fe/Mg \citep{Asplund2005}.

Terrestrial planets are separated into two major interior reservoirs with distinct compositions: a silicate mantle and an iron core. Most Si and Mg is present in the mantle, while Fe is partitioned between the mantle and core; hence, stellar Fe/Mg cannot be used directly as a proxy for the bulk mantle Fe/Mg. We chose mantle iron contents based on planet-formation models for the Solar nebula, which finds mantle iron contents between 6-20 wt\%, depending on oxygen fugacity \citep{Rubie2015,Fischer2017}. However, the Sun has a high abundance of iron compared to other stars in the Solar neighborhood (Fig.\ \ref{fig:Hypatia_comp}), so we also consider cases with less iron in the mantle, similar to other studies of terrestrial exoplanet interiors \citep[e.g.,][]{Wordsworth2018}. We consider iron contents ranging from 1 wt\% mantle FeO (i.e., low mantle oxygen fugacity with most of the iron in the core) to 20 wt\% mantle FeO in the mantle (i.e., high mantle oxygen fugacity). We assume that sufficient oxygen is available to oxidize all Mg, Si, and the appropriate amount of Fe. Given that the stars in the Hypatia catalog have a significantly higher oxygen abundance than Fe, Mg, and Si abundances, and that the C/O-ratio is low, oxygen should be sufficiently abundant to meet this assumption \citep{Bond2010,Hypatia,Hinkel2018}. 

We focus on the bulk composition of the silicate mantle since the mantle shares an interface with the atmosphere (the planetary surface). The composition of the core is not considered, including the presence of light elements such as Si, since the core only indirectly affects the evolution of the atmosphere \citep{Foley2016}. Including Si in the core would increase the mantle Mg/Si, however. Additionally, we assume the same core size as Earth for our simulated planets, to separate the effect of mantle composition from the effect of core size on planetary evolution. This approach does not yield a 1:1 relationship between the mantle Fe/Mg and bulk-planet Fe/Mg (see Table \ref{tab:comps}). For example, assuming an Earth-sized core, high iron content and low Mg/Si in the mantle yields bulk-planet Fe/Mg that is higher than would otherwise be realistic according to the Hypatia catalog. While these planets are likely to have a smaller core than we assume here, we generally expect core size to have only minor effects on our results (see Sect.\ \ref{ssec:Metho_Interior} and \ref{ssec:Disc_stablestrat}).

\begin{table}[h]
\centering
\caption{Range of compositions considered in this work: FeO mantle content (wt\%); molar Mg/Si ratio, corresponding to the 1$\sigma$ and 2$\sigma$ ranges of abundances in the Hypatia catalog; molar mantle Fe/Mg; molar bulk planet Fe/Mg. Compositions are indicated in Figures \ref{fig:Triangle_results} and \ref{fig:Triangle_comparison}. }\label{tab:comps}
\begin{tabular}{| c | c | c | c | c |} 
\hline 
FeO$_m$ & Mg/Si & Fe/Mg$_m$ & Fe/Mg$_p$ \\
\hline
1 & 0.72 (2$\sigma$) & 0.017 & 0.93 \\
1 & 0.89 (1$\sigma$) & 0.015 & 0.81 \\
1 & 1.37 (1$\sigma$) & 0.012 & 0.64 \\
1 & 1.71 (2$\sigma$) & 0.011 & 0.57 \\
20 & 0.72 (2$\sigma$) & 0.43 & 1.57 \\
20 & 0.89 (1$\sigma$) & 0.37 & 1.36 \\
20 & 1.37 (1$\sigma$) & 0.29 & 1.06 \\
20 & 1.71 (2$\sigma$) & 0.26 & 0.95 \\
\hline
\end{tabular}
\end{table}

\subsection{Magma ocean crystallization and cumulate overturn} \label{ssec:Metho_MO_cryst}
Our starting point is to consider the crystallization of the final global, deep, and compositionally homogeneous magma ocean (MO) above a completely formed metallic core. We made this choice because deep, global MOs can form late in the formation process of a terrestrial planet by giant impacts \citep[e.g.,][]{Nakajima2015} and tend to erase most heterogeneity that may have formed due to the crystallization of previous MOs \citep{Solomatov2000}. For Earth-sized planets, an MO may crystallize from the bottom--up \citep{Abe1997,Solomatov2000,Elkins2008,Ballmer2017_MO} or middle--out \citep{Labrosse2007,Stixrude2009,Nomura2011}. As melting curves (solidus and liquidus) for chondritic material \citep{Andrault2011} are usually estimated to be steeper than the silicate melt adiabat \citep{Wolf2018}, we consider bottom--up crystallization. 

We consider the fractional crystallization end-member scenario, rather than batch crystallization. For fractional crystallization, crystals form in the MO due to cooling, and sink to settle at the crystallization front \citep{Solomatov2015}. This allows for distinct compositional evolution (i.e., fractionation) of the MO liquid and forming crystals \citep[e.g.][]{Boukare2015}. For batch crystallization, rapid cooling of the MO prevents crystal settling such that co-existing liquid and crystals are instead frozen in place, thereby preventing distinct compositional reservoirs from forming \citep{Solomatov1993}. Since the outgassing of a thick steam atmosphere early in the lifetime of an MO has an efficient blanketing effect that slows down MO freezing, fractional crystallization is the more realistic scenario throughout much of the cooling sequence \citep{Elkins2008,Lebrun2013,Solomatov2015}. 

Due to fractional crystallization, the liquid MO becomes progressively enriched in FeO, as solids tend to incorporate MgO and SiO$_2$ over FeO \citep{Fiquet2010,Nomura2011,Andrault2012,Tateno2014,Boukare2015,Ballmer2017_MO}. We simulate this chemical evolution by calculating the composition of the solids that are in equilibrium with the (purely liquid) MO and extracting them from the liquid. We assume that the MO remains chemically and thermally homogeneous due to vigorous and pervasive turbulent convection. Crystal composition in the lower mantle is determined based on phase data of the MgO-FeO-SiO$_2$-system \citep{Boukare2015}, which considers the lower mantle phases (Mg,Fe)O (ferropericlase, fp), (Mg,Fe)SiO$_3$ (bridgmanite, bm), and SiO$_2$ (stishovite, st). The composition of fp and bm is determined using partition coefficients
\begin{linenomath*}
\begin{equation}
    K_{bm} = \frac{\mathrm{Fe}_{bm}/\mathrm{Mg}_{bm}}{\mathrm{Fe}_l/\mathrm{Mg}_l}\ \ \text{and} \ \ K_{fp} = \frac{\mathrm{Fe}_{fp}/\mathrm{Mg}_{fp}}{\mathrm{Fe}_l/\mathrm{Mg}_l},
    \label{eq:Kfpbm}
\end{equation}
\end{linenomath*}
for molar abundances of Fe and Mg in bm, fp and liquid (l), respectively. When both compounds exist at equilibrium, iron is partitioned according to $K_D = K_{bm}/K_{fp}$. We set constant values of $K_{bm} = 0.1$ and $K_{fp} = 0.9$ \citep{Tange2009,Boukare2015}. 

We assume that the composition of the upper mantle ($\leq$25 GPa) cumulates (material that crystallizes from the magma ocean) is identical to that of the upper mantle MO. This is motivated by the expected occurrence of small-scale overturns in the upper mantle, which are driven by unstable stratification. The timescale over which overturns occur depends on the viscosity of the solid \citep{Ballmer2017_MO, Maurice2017, Boukare2018} and lower mantle viscosity is much higher than upper mantle viscosity. The resulting difference in timescales leads to many overturns in the upper mantle before the occurrence of a final whole-mantle overturn. While the former sustains efficient mixing across the upper mantle, the latter is not expected to sustain efficient mixing across the whole mantle due to its large-scale nature. Overturning instigates decompressional melting and re-melting, thereby homogenizing the upper mantle material and, thus, justifying our assumption that the solid and the melt have the same composition in the upper mantle.

A whole-mantle overturn is promoted by unstable stratification between FeO-rich upper mantle cumulates and FeO-poor lower mantle materials \citep{Elkins2008}. Such a large-scale overturn leads to the sinking of FeO-rich material through the mantle to form a compositionally distinct layer at the base of the mantle \citep{Hess2001}. We model the whole-mantle overturn event by flipping the compositional profile from the MO crystallization upside-down. We then calculate the resulting density profile using 
\begin{linenomath*}
\begin{equation} \label{eq:Fe_Density}
\Delta \rho_{comp} = 850 \cdot \text{FeO}/(\text{FeO}+\text{MgO}),    
\end{equation}
\end{linenomath*}
for a density contrast of 850 kg/m$^3$ between Mg- and Fe-silicate end-members \citep[e.g.,][]{Wolf2015}. If $\Delta \rho_{comp}$ exceeds a critical value $\Delta \rho_{crit}$, the layer resists entrainment and mixing by solid-state convection, and stratification remains stable for most of the planet's lifetime \citep{Kellogg1999,Tosi2013,Plesa2014}. Otherwise, the neighboring layers are mixed to form a homogeneous layer. We iterate through the layers and check that they are stable based on our stability criterion. In the literature, the stability criterion is usually given in terms of a buoyancy number, $\mathcal{B} = \Delta \rho / \rho \alpha T_{CMB}$, where $\rho=4000$kg m$^{-3}$, $\alpha = 1.5 \cdot 10^{-5}$K$^{-1}$, and $T_{CMB}=4000$K. As a reference value, we consider $\Delta \rho_{crit} = 225$kg/m$^3$, which corresponds to $\mathcal{B} \approx 0.94$ \citep{Kellogg1999}. This is a conservative criterion, compared to other literature values \citep[e.g.,][]{Tosi2013,Li2014,Ballmer2017_MO}. We explore the effect of the value of this parameter on the occurrence of stable stratification, given our compositional spread, in Section \ref{ssec:Disc_stablestrat}.

\subsection{Long-term interior evolution} \label{ssec:Metho_Interior}
The structure and mineralogy of the mantle resulting from MO crystallization and cumulate overturn models provide the initial conditions for the subsequent long-term evolution of the interior. We model the long-term thermal evolution of the mantle using the 1D parameterized convection code SPIDER \citep{Bower2018}. SPIDER solves the energy balance in a planetary mantle by using mixing length theory (MLT). The long-term thermal evolution predicted by MLT models agrees with the predictions of 3-D convection simulations \citep{Kamata2018,Wagner2019}. Adopting the formulation from \cite{Kamata2018}, we describe the mixing length profile using $a=0.75$ and $b=0.45$ \citep[Eqs.~13--14 \& Fig.~2,][]{Kamata2018}. We neglect the removal of latent heat by melt formation and extraction.

\begin{figure}
   \centering
   \includegraphics[width=\hsize]{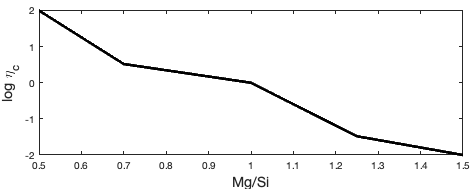} 
      \caption{Compositional correction of viscosity, $\eta_c$, as a function of molar mantle Mg/Si.}
      \label{fig:Viscosity_comp}
      \vspace{-15pt}
\end{figure}

We use a viscosity law with pressure-dependent activation volume $V_a(P)$ \citep{Yamazaki2001} and a composition-dependent prefactor $\eta_c$,
\begin{linenomath*}
\begin{equation}
    \eta (T,P) = \eta_0 \eta_c \exp\left( \frac{E_a + V_a(P) P}{R T} - \frac{E_a + V_a(P_0) P_0}{R T_0} \right),
\end{equation}
\end{linenomath*}
where $V_a(P) = V_{a,0}\exp(-P/P_{scale})$, for reference viscosity $\eta_0$ at $P_0$ and $T_0$, and activation energy $E_a$. The compositional term $\eta_c$ depends solely on the Mg/Si-ratio of the mantle. It decreases with Mg/Si from $\eta_c \approx 10$ at Mg/Si=0.5, which corresponds to a st-rich composition \citep{Xu2017}, to $\eta_c = 1$ at Mg/Si = 1.0, and, finally, to $\eta_c = 10^{-2}$ for Mg/Si $\geq$ 1.5, which corresponds to a fp-rich composition \citep{Ballmer2017_silica} (Fig.~\ref{fig:Viscosity_comp}). We implemented a viscosity cut-off for the viscous upper boundary layer at $\eta = 10^{27}$ Pa$\cdot$s. Other viscosity parameters were calibrated to ensure the viscosity profile adhered to data-derived estimates for the viscosity structure of Earth's mantle \citep[e.g.,][]{Forte1996,Mitrovica2004,Paulson2005}: $\eta_0 = 10^{20.5}$ Pa$\cdot$s for $T_0 = 3500$ K and $P_0 = 140$ GPa, $E_a = 175$ kJ/mol, $V_{a,0} = 4.0\cdot 10^{-6}$ J/mol/Pa, and $P_{scale} = 200$ GPa.

The possible occurrence of stable stratification as constrained by our overturn model is imposed in the evolution model by introducing a mid-mantle interface. This interface is implemented to separate two convecting domains of different composition by adjusting the mixing length profile. The mixing-length profile is rescaled (with the given $a$ and $b$ parameters) to account for the different thicknesses of each layer. The mixing length is zero at the mid-mantle interface, which enforces a thermal boundary layer for the duration of the evolution (10 Gyr). Importantly, the transport (viscosity) and material (density) properties of each domain are set independently according to the compositional profile that results from MO crystallization and cumulate overturn.

The long-term evolution of planetary mantles is influenced by the dynamics of the lithosphere. The two end-member cooling regimes involve a mobile lithosphere, which participates directly in mantle convection (plate tectonics, PT), and an immobile lithosphere that hinders efficient cooling (stagnant lid, SL). Therefore, we consider a fast-cooling (PT) and a slow-cooling (SL) end-member. For the SL case, we impose an additional viscous lid (two orders of magnitude more viscous than usual) at the top of the mantle. This lid efficiently restricts cooling of the mantle to mimic the behavior of a stagnant lid. The thickness of the lid is 200 km, which is consistent with estimates for Earth-sized SL planets \citep[see][and references therein]{Reese1999}. In reality, the lid thickness depends on mantle temperatures, but the dependence is not well-constrained. Imposing a constant lid thickness implies that the lid formed relatively early and with a fairly constant thickness over time. For the PT-case the viscosity structure is unaltered. This is because the mixing length theory that we adopted naturally extends the thermal boundary layer to the surface, effectively assuming that the full depth of the domain is participating in convection. The dynamic regime of the lithosphere is imposed throughout model evolution. It is not known whether bulk composition plays a role to stabilize the PT versus\ the SL regime or to promote transitions between regimes (see Sect.\ \ref{sec:Disc_Regime}). Regime changes are predicted by geodynamic simulations, and consistent with the geologic record for Earth and Venus \citep[e.g.,][]{Sleep2000,Armann2012,Korenaga2013}. Here we take the simplified approach to consider end-member scenarios, assuming that a planet remains locked in either the PT or the SL regime.

The surface temperature is set to 273 K, while the core-mantle boundary temperature (initially 4200 K) evolves according to a core cooling model; the core has a thermal energy budget that decreases as heat is removed by convection in the overlying mantle. We keep core size constant in this model, as it would have a minor effect on thermal evolution compared to the composition-dependent viscosity correction we implement here. The model also includes radiogenic heating of the mantle by long-lived isotopes which decay over time ($^{40}$K, $^{232}$Th, $^{235}$U and $^{238}$U), at Earth-like initial concentrations \citep{Ruedas2017}. A constant conductivity of 4 W/m/K is included in the models, which has a negligible effect on heat transport compared to convection \citep{Bower2018}.

\subsection{Melting, outgassing, and ingassing} \label{ssec:Metho_Volatiles}
We modeled the volatile evolution of the interior and secondary atmosphere with a box model, where volatile exchange between reservoirs consists of outgassing (transport from the interior to the atmosphere) and ingassing (transport from the atmosphere to the interior). We considered three volatile reservoirs in our model: the mantle, atmosphere, and lithosphere. For the SL regime, we also refer to the lithosphere reservoir as the lid reservoir. We focus exclusively on water since it partitions similarly to most other volatiles (excluding carbon\footnote{Carbon is stored as graphite or diamond, or as metallic carbides under reducing conditions, and these phases do not readily partition into silicate melt, unlike other volatile phases (which are, therefore, considered incompatible with the solid)}; however, carbon stored in fluids or as carbonates is incompatible; \cite{Dasgupta2010}. Thus, water has a pronounced influence on melt generation by depressing the melting curve (solidus). We included three Earth ocean masses of water \citep{Bercovici2003,Houser2016}, initially distributed between the atmosphere (1 Earth ocean mass) and interior (2 Earth ocean masses).

Outgassing is driven by volcanism, which, in turn, is driven by melt generation in the interior. Outgassing rates are determined by calculating melt formation and melt volatile content. Following \cite{Katz2003}, the local melt fraction $\phi$ at each depth (pressure) is calculated as a function of temperature $T$,
\begin{linenomath*}
\begin{equation}
    \phi(P,T) = \left( \frac{T - T_{sol}(P)}{T_{liq}(P) - T_{sol}(P)} \right)^{1.5},
    \label{eq:Melt_fract}
\end{equation}
\end{linenomath*}
with solidus and liquidus profiles, $T_{sol}(P)$ and $T_{liq}(P),$ taken from \cite{Andrault2011}. Because these profiles are constructed for chondritic composition, we applied a correction to the solidus profile as a function of iron content \citep{Dorn2018_outgassing} since iron depresses the solidus temperature. The effects of Mg/Si on the solidus are relatively unexplored (for current efforts, see e.g.,\ \cite{Brugman2020}) and comparisons of melting experiments for chondritic (Mg/Si=1.05) and peridotitic (Mg/Si=1.26) compositions reveal that there is a negligible effect of Mg/Si on the solidus in this compositional range at pressures <5 GPa \citep{Andrault2018}. We do not correct for composition in terms of Mg/Si as the effect is much smaller for the compositional range of interest \citep{deKoker2013,Boukare2015}.

Melt generation usually occurs at the top of hot plumes (i.e., upwellings in the mantle), and at locations where the crust is very thin and mantle material approaches the surface, allowing decompression melting to occur. In order to address lateral thermal variations in the mantle in our 1D model, we considered a Gaussian distribution with a standard deviation of 125 K that is centered on the 1D temperature profile. This value of 125 K corresponds to half the estimated excess temperature in plumes \citep{Schilling1991}. Bulk melt fraction as a function of depth is obtained by integrating Eq.\ \ref{eq:Melt_fract} with the Gaussian temperature distribution. The total melt fraction is determined each time step, and transports volatiles (i.e., water) from the interior to the surface where they are outgassed.

We describe the water content in the melt with equilibrium partitioning between solid and liquid because water behaves like an incompatible element and, thus, preferentially partitions into the melt:
\begin{linenomath*}
\begin{equation}
    X_w = \frac{X_{w,bulk}}{D + \phi (1 - D)}
    \label{eq:Melt_water_cont}
\end{equation}
\end{linenomath*}
for the molar water concentration in the melt phase ($X_w$), molar water concentration in the bulk material ($X_{w,bulk}$), and partition coefficient $D$ \citep{Zou1998,Katz2003}, where $D=0.01$ for water \citep{Aubaud2004}. The melting behavior of mantle material depends on the concentration of water in the melt \citep{Katz2003} since the presence of water depresses the solidus. Therefore, the melt fraction and water concentration in the melt are mutually dependent. In order to compute melt volume, this non-linearity requires that we iteratively solve Equations \ref{eq:Melt_fract} and \ref{eq:Melt_water_cont}, with the solidus depression by water given by \cite{Katz2003}, until equilibrium is reached. The solidus depression by water we use here is parametrized for peridotitic composition since no parametrization exists for other compositions.

The outgassing rate is calculated by determining the melt volume and water concentration and multiplying with the extrusion efficiency. The transfer of volatiles from the interior to the atmosphere is only possible when melt extrudes (i.e., reaches the surface and is not trapped in the lithosphere). In the SL regime, volatiles trapped within intrusive magmas are stored in the lid reservoir. Planets in the SL regime have lower melt extrusion rates due to the large thickness of the lithosphere, compared to the PT regime. We assume an intrusive-to-extrusive ratio of 9:1 for planets in the SL regime \citep{Crisp1984}, based on observations of Earth. Additionally, we assume that all melt, wherever it is generated in the mantle, is transported to the near-surface and can source either intrusive or extrusive magmatism. We assume that outgassing of the lithosphere reservoir, which requires melting of crustal material, does not occur.

On planets in the SL regime, ingassing can occur by lithospheric delamination, where blobs of crustal material sink into the mantle from the base of the stagnant lid, transporting volatiles from the lid to the mantle reservoir. For lid thickness to remain constant over time, the delamination flux must be balanced by mantle melt production. Delaminating material originates from the bottom of the lid, where the material is hot because of the proximity to the hot upper mantle. Therefore, hydrous minerals are not stable \citep{Frost2006,Iwamori2007}, such that water is exclusively transported by nominally anhydrous minerals. In our model, delaminating material has the same concentration of water as the lid, with a maximum concentration of 400 ppmwt \citep{Ferot2012} to account for the maximum capacity of nominally anhydrous minerals. In contrast to ingassing from the lid reservoir, volatile transport from the atmosphere to the mantle is inefficient for planets with an immobile and sufficiently thick lid \citep{Foley2018}. The atmospheric ingassing rate in the SL case is, therefore, set to zero.

In the PT regime, we combine the lithosphere and atmosphere reservoirs when calculating the ingassing rate. This inherently sets the extrusion efficiency to unity, and allows us to model mantle ingassing by subduction as direct transport of volatiles from the atmosphere to the mantle reservoir. We merge these reservoirs because ingassed volatiles originate from both the lithosphere and atmosphere reservoirs. These reservoirs interact by hydrothermal alterations in the oceanic crust, which transports volatiles from the atmosphere to lithosphere reservoir \citep{Wolery1976}. Ocean-lithosphere fluxes are not well constrained and depend on local chemistry, and are therefore not included. Our chosen Earth-like surface temperature of 273 K promotes the stabilization of liquid oceans and submerged oceanic plates, thereby allowing metamorphic processes to incorporate water into the lithosphere (as on Earth).

During subduction, only water that is released beyond a certain depth reaches the mantle, while water released at shallower depths causes arc volcanism and directly returns to the atmosphere reservoir. Water retention (and therefore the ingassing rate) depends on the stability of hydrous minerals, which, in turn, is affected by composition and mantle temperature. However, hydrous phase stability mainly depends on composition in terms of elements that we do not consider here \citep{Frost2006,Iwamori2007}. Additionally, the complex feedback mechanisms between mantle temperature, melting, and tectonics would require high-resolution 2D or even 3D models \citep{Rolf2011,Lourencco2016}. Therefore, in light of the nature of our models, which are designed to investigate first-order effects only, we assume a constant PT ingassing rate. We assume an ingassing rate of $2 \cdot 10^{11}$ kg/yr for water. This is the estimated water transport rate by subducting slabs beyond a depth of 250 km for Earth \citep{vanKeken2011}.

\section{Results}\label{sec:ResDis}
To address the long-term evolution of terrestrial planets, we explore a coupled interior-atmosphere evolution model as a function of bulk planetary composition, namely, MgO-FeO-SiO$_2$ (Fig.~\ref{fig:Hypatia_comp}). Our modeled planets have the same mass, radius, and core size as Earth. The initial condition of the mantle is calculated from a simplified MO crystallization model (Sec.\ \ref{ssec:Res_MOcryst}). Thereafter, a long-term evolution model determines the thermal structure of the interior and a volatile exchange model determines how water migrates between the interior and atmosphere (Sec.\ \ref{ssec:Res_InteriorEvo}).

\subsection{Magma ocean crystallization and cumulate overturn} \label{ssec:Res_MOcryst}
\begin{figure}[h!]
    \centering
    \includegraphics[width=\hsize]{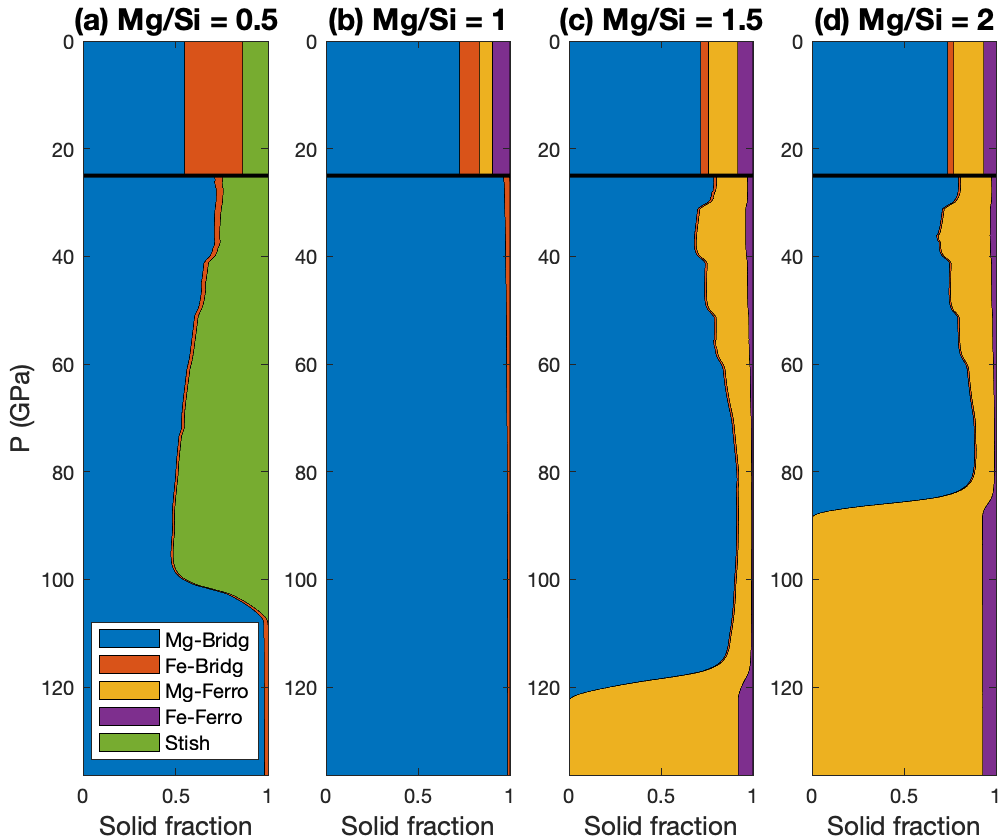}
    \caption{Composition of the crystallizing phase during magma ocean crystallization, in terms of minerals Mg/Fe-bridgmanite, Mg/Fe-ferropericlase, and stishovite, as a function of pressure (GPa). The depth of the mantle extends from the surface (0 GPa), to the core-mantle boundary (138 GPa, 2891 km). Cumulates are re-homogenized at pressures of 0-25 GPa (indicated by a black line) and, therefore, demonstrate constant composition (see Sect.\ \ref{ssec:Metho_MO_cryst}). Compositions are 8 wt\% FeO (Earth-like) and Mg/Si is annotated.}
    \label{fig:Earth_cryst}
\end{figure}

We model bottom-up crystallization of an MO for bulk silicate compositions ranging from Mg/Si=0.5 to Mg/Si=2.0, and for 1, 8, and 20 wt\% FeO. The predicted compositional profile of solid material that is crystallized from the MO depends mainly on the bulk Mg/Si-ratio. FeO content only marginally affects the mineralogy of the crystallizing solid. Bridgmanite (bm, (Mg,Fe)SiO$_3$) is the main mineral to crystallize in the lower mantle for a wide range of bulk compositions. In turn, stishovite (st, SiO$_2$) only appears for low Mg/Si-ratios (Fig.\ \ref{fig:Earth_cryst}a) and ferropericlase (fp, (Mg,Fe)O) only for high Mg/Si-ratios (Fig.\ \ref{fig:Earth_cryst}c,d). The compositional profile in the upper mantle (less than 25 GPa) is assumed to be homogeneous according to our assumption that small-scale overturns continuously homogenize the newly formed solid (Sect.\ \ref{ssec:Metho_MO_cryst}). The composition of the pre-overturn upper mantle strongly depends on the minerals that are removed from the MO during the fractional crystallization of the lower mantle (and how much FeO they incorporate), in addition to bulk composition. 

\begin{figure}
    \centering
    \includegraphics[width=\hsize]{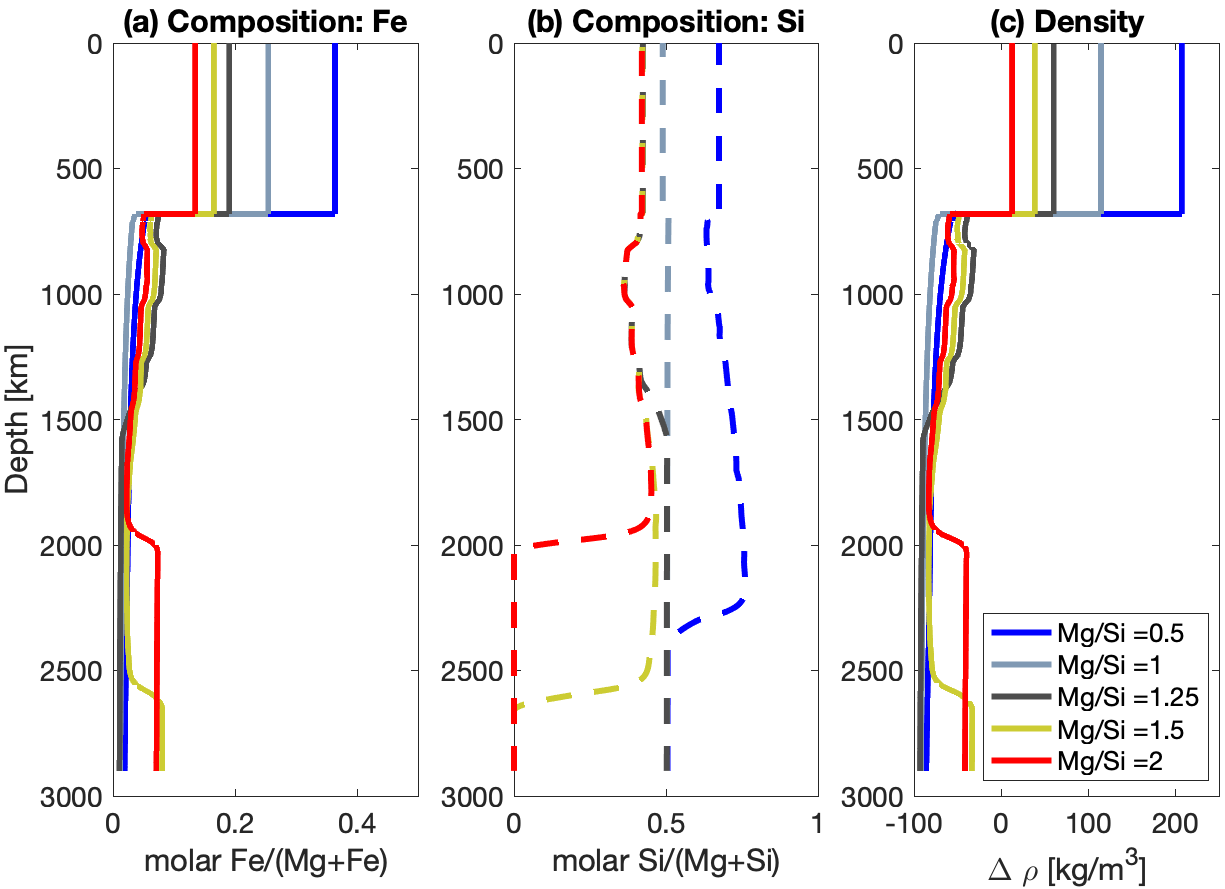}
    \caption{Predicted (a+b) compositional and (c) density profiles of the initial solid mantle (i.e., after MO crystallization and prior to overturn), for bulk planet compositions of 8 wt\% FeO and mantle molar Mg/Si ranging from 0.5 to 2.0. Compositional profiles are reported as molar ratios of Fe/(Mg+Fe) (a) and Si/(Mg+Si) (b). The density difference is between the crystallized solid, and pyrolite (i.e.\ Mg\# = 0.88; see Eq.\ \ref{eq:Fe_Density}). Crystallization proceeds from the bottom--up.}
    \label{fig:Composition_lines}
\end{figure}

In all cases, this upper part of the initial mantle is the most FeO-enriched part (Figure \ref{fig:Composition_lines}a) because iron preferentially partitions into the melt and the MO crystallizes in a bottom-up manner. Bulk Mg/Si is the main control on the degree of FeO enrichment of the upper mantle material, as it determines whether the lower mantle consists mainly of minerals with a high iron partition coefficient (fp) or a low iron partition coefficient (bm, st). Thus, the iron content in the upper mantle (Fig.\ \ref{fig:Composition_lines}a) and the related density contrast across the mantle (Fig.\ \ref{fig:Composition_lines}b) increases with decreasing bulk Mg/Si. In terms of Mg and Si, partitioning coefficients are close to unity for mantle minerals considered here and, hence, the compositional profile in terms of Si/(Mg+Si) is rather constant across most of the mantle (Fig.\ \ref{fig:Composition_lines}a). For Earth-like mantle compositions (8 wt\% FeO, molar Mg/Si=1.2, \cite{McDonough1995}), we predict a lower-mantle mineralogy that is dominated by bridgmanite and complemented by a minor amount of ferropericlase (Fig.\ \ref{fig:Earth_cryst}). This prediction is matched by our current knowledge of lower mantle mineralogy of Earth \citep{Kesson1998}.

\begin{figure}
    \centering
    \includegraphics[width=\hsize]{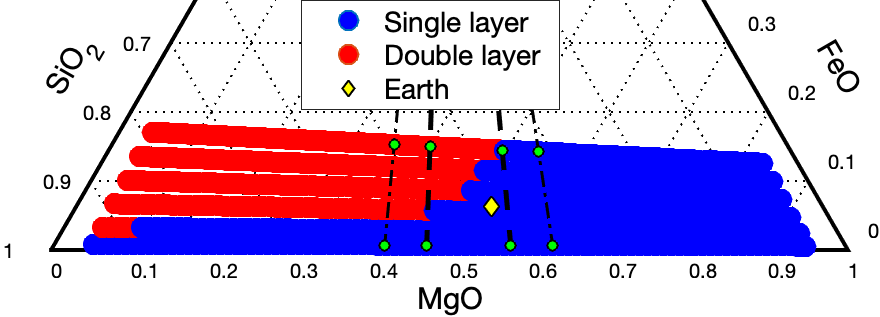}
    \caption{Predicted stratification of the mantle after overturn as a function of bulk mantle composition in terms of molar MgO, FeO, and SiO$_2$, assuming $\Delta \rho_{crit}=225$ kg m$^{-3}$. Blue and red dots denote cases with density differences across the mantle smaller than and larger than $\Delta \rho_{crit}$ (i.e., with a well-mixed and stably stratified mantle), respectively. Composition is shown for a range of bulk mantle Mg/Si from 0.05 to 20, and FeO from 1 wt\% to 20 wt\%. The 1$\sigma$ and 2$\sigma$ ranges of Mg/Si-ratios from the Hypatia catalog are shown as vertical dashed and dot-dashed lines, respectively. Green dots on these lines correspond to compositions shown in Table \ref{tab:comps}. The yellow diamond shows Earth's mantle composition \citep{McDonough2003}.} 
    \label{fig:Triangle_results}
\end{figure} 

All predicted profiles after MO crystallization are gravitationally unstable and, hence, they drive a mantle-scale overturn. This result is demonstrated in Figure \ref{fig:Composition_lines}b and corroborated by the expected unstable thermal stratification, which results from fractional crystallization of the MO \citep{Boukare2018}. After overturn, the propensity of the inverted density profile to whole-mantle convection or stable stratification depends on the total density contrast relative to a critical density difference, which we assume to be 225 kg m$^{-3}$ (but cf. Sect.\ 4.1).

Around 50\% of planets with compositions from the Hypatia catalog are predicted to result in rocky planets with stable stratification and, hence, persistent double-layered convection. We find that no large-scale stable stratification is formed for a planet with an Earth-like composition (Fig.\ \ref{fig:Triangle_results}), which is consistent with seismic constraints \citep[e.g.,][]{Helffrich2001}. Low bulk Mg/Si promotes iron enrichment in the upper mantle, and thereby stable stratification of the mantle for billions of years. High bulk FeO contents also promote stable stratification.

\subsection{Long-term evolution of interior and atmosphere} \label{ssec:Res_InteriorEvo}
\begin{figure*}[!h]
   \resizebox{0.99\hsize}{!}{\includegraphics{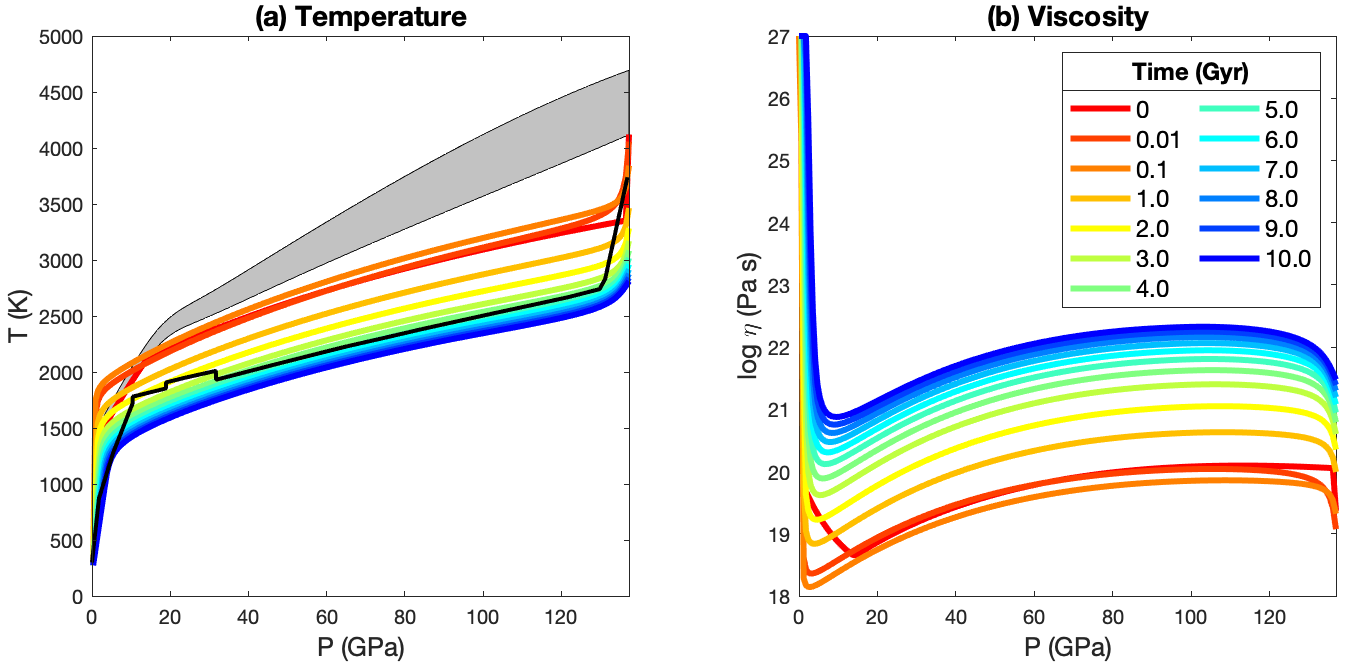}}
    \caption{Evolution of (a) mantle temperature and (b) viscosity for a terrestrial planet with Earth-like composition (Mg/Si=1.2, 8 wt\% FeO) as a function of pressure (GPa), ranging from the surface (0 GPa) to the core-mantle boundary (138 GPa, 2891 km). In panel (a), the solidus and liquidus from \cite{Andrault2011} are shown as thin black lines that bound a grey area; the temperature profile of the modern Earth is shown as a thick black line \citep{Stacey2008}.}
    \label{fig:Earth_comp_thermal_evo}
\end{figure*}

We investigate the long-term coupled evolution of the interior and atmosphere for a wide range of bulk compositions in the plate-tectonics (PT) and stagnant-lid (SL) dynamic regimes. As a reference case, we consider a terrestrial planet in the PT regime with a bulk mantle composition of (molar) Mg/Si = 1.2 and with 8 wt\% FeO (Fig.\ \ref{fig:Earth_comp_thermal_evo}), which corresponds approximately to that of the Earth \citep{Wang2018comp}. In this case, average interior temperatures decrease for most of the planet's life (Fig.~\ref{fig:Earth_comp_thermal_evo}a), mostly remaining below the solidus (except in the uppermost part at pressures $<$5 GPa). Mantle viscosity increases with time as temperature decreases (Fig.\ \ref{fig:Earth_comp_thermal_evo}b). Our thermal evolution model can roughly reproduce estimates for the current thermal profile of the Earth's mantle (Fig.\ \ref{fig:Earth_comp_thermal_evo}), except near the core-mantle boundary, where even these estimates remain uncertain. Since our viscosity profiles have been calibrated based on our current understanding of Earth mantle viscosity structure, there is naturally a good match with estimates from geoid inversions (e.g., \citep{Forte1996}). However, we did not consider viscosity jumps near phase transitions.

\begin{figure*}
   \resizebox{\hsize}{!}{\includegraphics{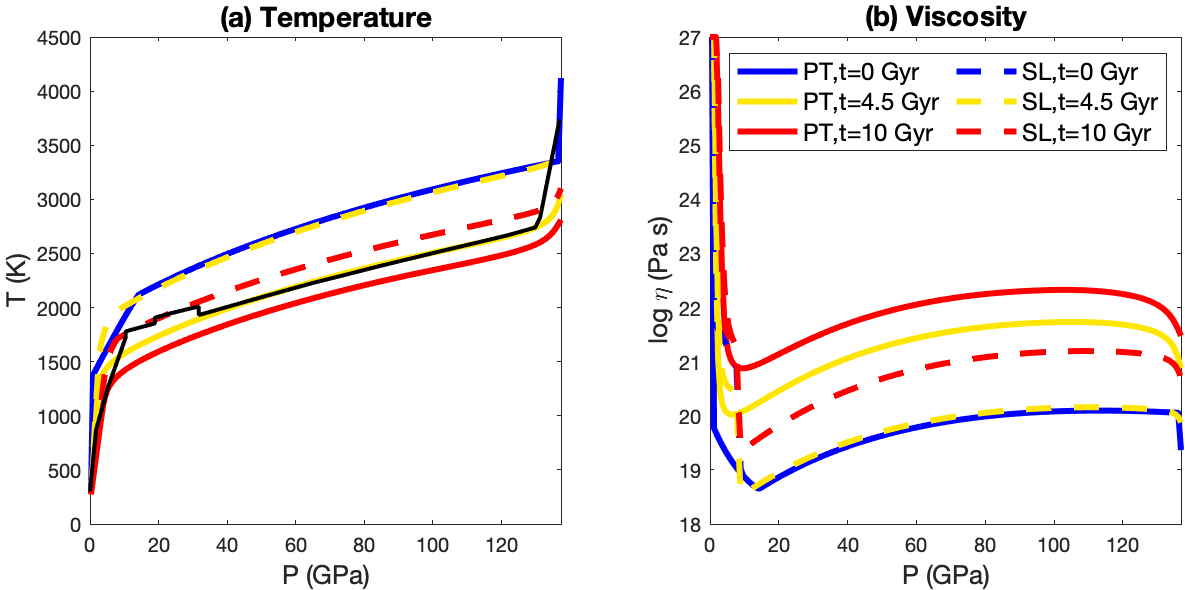}}
   \caption{Evolution of the mantle temperature (a) and viscosity (b) for a planet with Earth-like composition (Mg/Si=1.2, 8 wt\% FeO) in the PT- and SL-cases  as a function of pressure (GPa), ranging from the surface (0 GPa) to the core-mantle boundary (138 GPa, 2891 km). Profiles are shown for the initial condition (t=0 Gyr), current age of the Earth (t=4.55 Gyr), and for the estimated lifetime of the Sun (t=10.0 Gyr). Independently estimated profiles for the present-day Earth are shown as black lines \citep{Stacey2008}.}
   \label{fig:Earth_regimes}
\end{figure*}

Next, we compared the reference case with an SL-case for the same bulk composition (Fig.\ \ref{fig:Earth_regimes}). We find that a planet in the SL-case cools systematically slower than in the PT-case, displaying a higher average mantle temperature at any given age (time). This delayed cooling is explained by the presence of a viscous lid in the uppermost 200 km of the interior, which slows down cooling of the interior (Fig.\ \ref{fig:Earth_regimes}a). Accordingly, the SL-case reaches temperatures similar to Earth's present-day (i.e., 4.5 Gyr) thermal profile only after $\sim$10 Gyr.

\begin{figure*}
   \resizebox{\hsize}{!}{\includegraphics{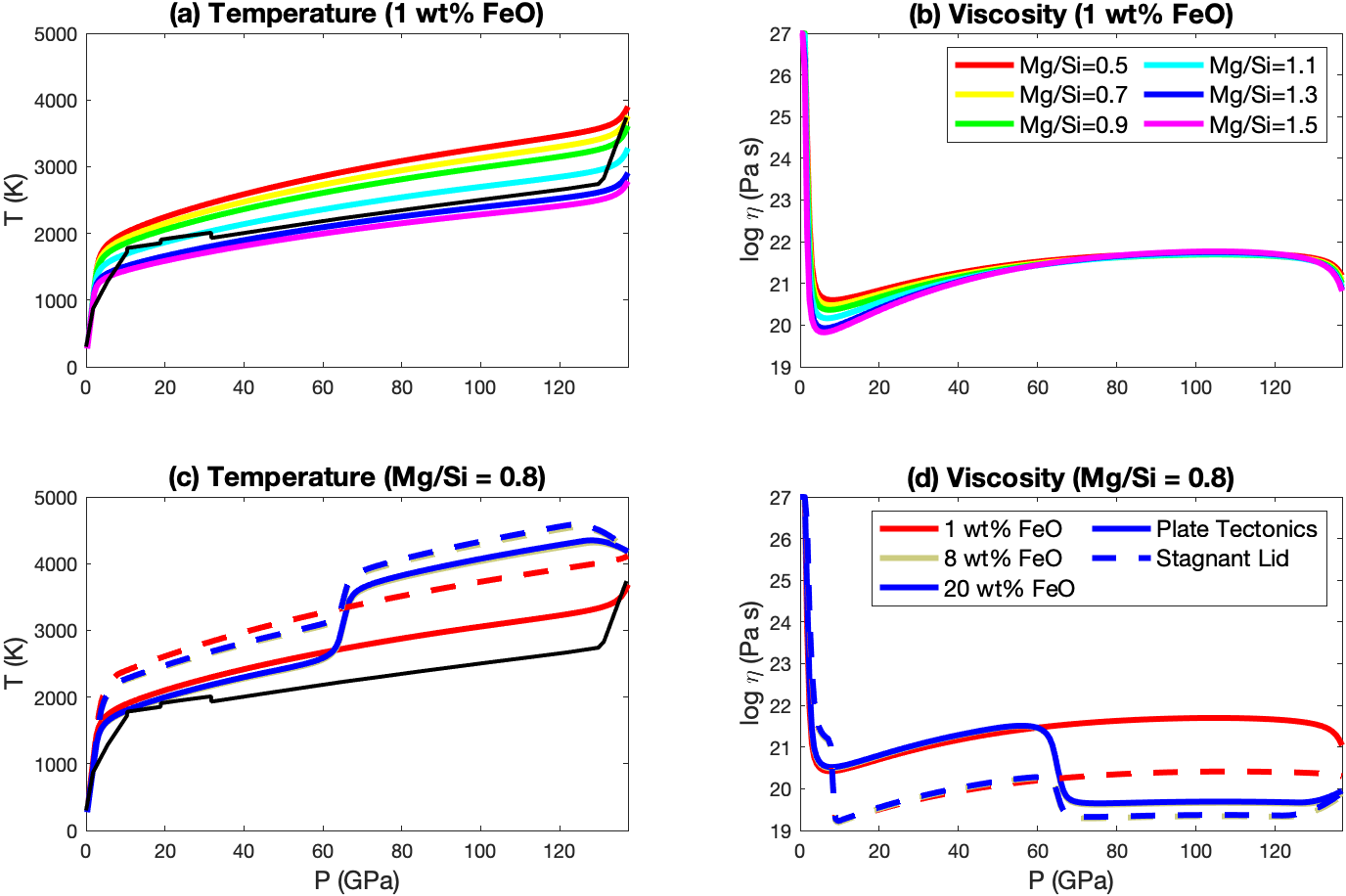}}
   \caption{Temperature (left column of panels) and viscosity (right column of panels) profiles at t = 4.55 Gyr for planets with variable bulk-mantle Mg/Si and FeO content as annotated, as a function of pressure (GPa), ranging from the surface (0 GPa) to the core-mantle boundary (138 GPa, 2891 km). In the top row of panels, FeO content is 1 wt\%. In the bottom row of panels, Mg/Si = 0.8. Profiles are shown for t = 4.55 Gyr. PT- and SL-cases are shown as solid and dashed lines, respectively. The cases shown in the bottom row with 8 and 20 wt\% FeO demonstrate stably stratified mantles.}
   \label{fig:Therm_comp}
\end{figure*}

\subsubsection{Influence of bulk mantle composition}
We find that the thermal evolution of the interior is strongly affected by the bulk Mg/Si-ratio of the planet. Planets with high Mg/Si-ratios tend to cool faster than planets with low Mg/Si-ratios (Fig.\ \ref{fig:Therm_comp}a). This is explained by the higher viscosity of low Mg/Si-cases, which slows down heat transport through the mantle and to the surface. Furthermore, planets with low Mg/Si-ratios display slightly steeper temperature profiles than planets with high Mg/Si. Planetary materials with Mg/Si < 0.5 and Mg/Si > 1.5 have a viscosity differing by up to three orders of magnitude for a given convecting temperature profile, but the slower cooling of planets with low Mg/Si effectively reduces this viscosity variation (Fig.\ \ref{fig:Therm_comp}b). 

Mantle layering also has a strong influence on planetary interior evolution, and as mentioned above, the propensity to mantle layering itself is a function of bulk composition, including Mg/Si. For identical compositions and dynamic regimes, planets with layered mantles display much hotter lower mantles than planets without stratification, and slightly cooler upper layers (Fig.~\ref{fig:Therm_comp}c). Mantle viscosities mirror this trend, with the lower mantles having a viscosity of about 1.5 orders of magnitude lower than for planets without stable layering (Fig.\ \ref{fig:Therm_comp}d). This prediction is because the thermal boundary layer at the compositional boundary restricts the cooling rate of the lowermost mantle. As shown in Section 3.1, the two layers typically have distinct Mg/Si-ratios, with the upper layer generally having a higher Mg/Si than the lower layer. However, the Mg/Si-contrast between layers is small, and the associated change in viscosity has a very limited effect on the long-term interior evolution. Finally, within the regime of stable mantle stratification, there is little to no difference between cases with 8 and 20 wt\% FeO in terms of long-term thermal evolution, demonstrating again that Mg/Si is the stronger control on the evolution of planetary mantles.

\begin{figure*}
   \resizebox{\hsize}{!}{\includegraphics{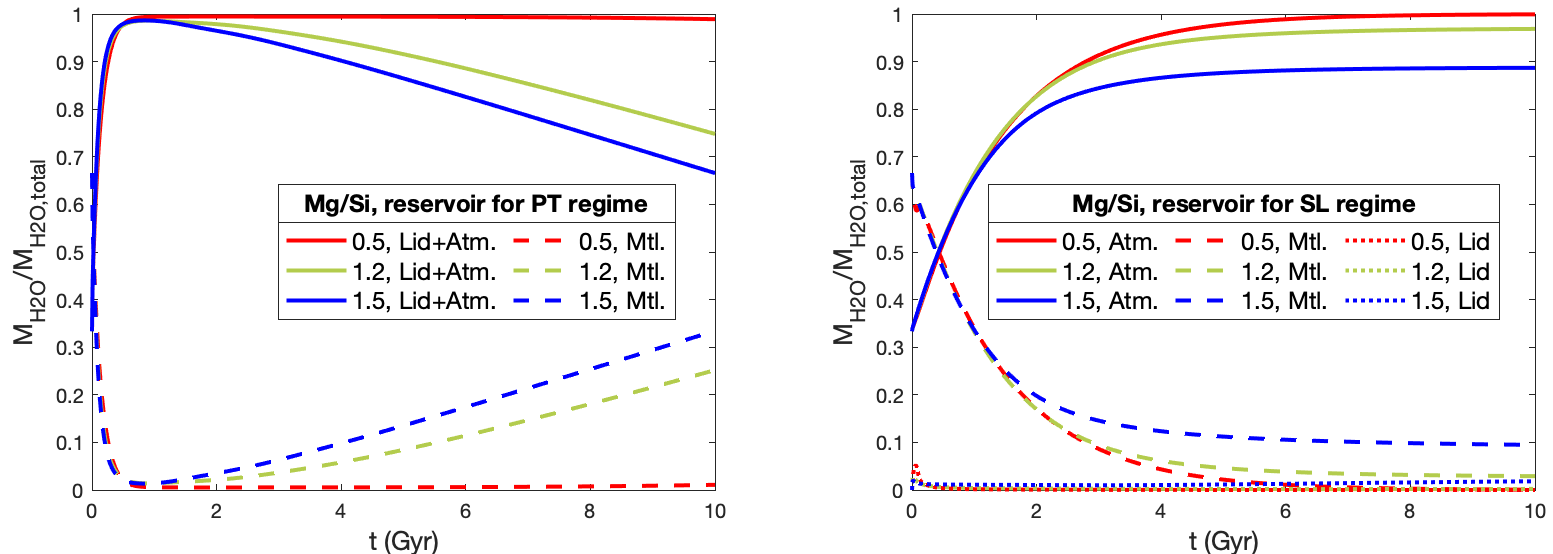}}
      \caption{Evolution of volatile reservoirs (mantle - dashed line, atmosphere - solid line, lid - dotted line) of a planet in the plate tectonics (left) and stagnant lid (right) regime, with Earth-like bulk mantle iron (8 wt\% FeO), as a function of model time (Gyr). Total water mass $M_{H2O,total}$ is equal to three Earth ocean masses. Note that the lid reservoir occurs only in the SL regime, being fed by intrusions. In the PT regime, the lid reservoir is counted towards the atmosphere reservoir, and eventually becomes part of the ingassing term (see Sect. 2).}
      \label{fig:Res_Vol}
\end{figure*}

\subsubsection{Volatile fluxes}
Based on the temperature profiles predicted by our thermal evolution model, we computed the fluxes of water between the secondary atmosphere, mantle, and lid reservoirs for a bulk planet water content of three Earth ocean masses (Fig.\ \ref{fig:Res_Vol}). Planets in the PT regime are characterized by efficient outgassing in the first Gyr, which brings more than 95\% of the planets water budget into the atmosphere, independent of composition. After about 1 Gyr, the net ingassing of the mantle becomes dominant for planets with high Mg/Si. In contrast, planets with low Mg/Si demonstrate net outgassing until 5 Gyr. This prediction is explained well by the higher average temperature of planets with low Mg/Si (Fig.~\ref{fig:Therm_comp}), leading to greater melt production over the lifetime of the planet. In terms of the Earth's volatile reservoirs, while we do not accurately reproduce common estimates for the mantle reservoir (e.g.,\ 3-6 Earth Ocean masses, \cite{Bercovici2003}), the water inventory of Earth's mantle remains a matter of debate. Plausible interpretations are discussed in Section \ref{ssec:Disc_volatiles}. 

In the SL regime, planetary volatile reservoirs and fluxes follow a different evolutionary path than in the PT regime. SL-cases also start with a phase of strong net outgassing, but outgassing is slower than for the PT regime, and we assumed that no ingassing from the atmosphere to the mantle can occur in this regime. Outgassing continues for at least 4 Gyr, with low-Mg/Si cases experiencing continued outgassing up to 8 Gyr. The volatile reservoirs approach a steady state, in which most volatiles reside in the atmosphere. The mantle is stripped of virtually all its volatiles for low Mg/Si-cases, while for high Mg/Si cases, it retains a small but steadily decreasing amount of volatiles. The lid reservoir does not contain more than a few percent of the bulk volatile budget because of high delamination fluxes as long as outgassing is efficient.

\section{Discussion} \label{sec:Disc}
\subsection{Stable stratification}
\label{ssec:Disc_stablestrat}
We find planetary interiors with and without stable mantle stratification for the range of planetary bulk compositions constrained by the Hypatia catalog (Fig.\ \ref{fig:Triangle_results}). This result implies that around half of Earth-sized exoplanets may have a stratified mantle with important implications for their long-term evolution. Furthermore, for planets with stratified mantles, the upper mantle composition is different from that of the bulk mantle; hence, the bulk mantle composition is no longer representative of the composition of the volatile-carrying reservoir. Since mineralogy affects the melting behavior of the mantle, stratification thus affects the evolution of volatile reservoirs. Stable mantle stratification is more likely for planets with a low bulk Mg/Si-ratio, and with high mantle iron content. Additionally, stable stratification is more likely to develop for planets with high internal oxygen fugacity and therefore small core size. Thus, the presence or absence of mantle layering can help constrain core size.

Depending on the intrinsic density and viscosity contrast between the two compositionally distinct layers, the boundary between them can develop significant topography due to convective stresses \citep{Kellogg1999}. Likewise, the entrainment of material from one layer by convection in the other layer may occur, thereby diluting the density contrast that is maintaining the stratification. However, for high initial density contrasts, the entrainment rate is sufficiently low, and stable stratification persists throughout the lifetime of a mature planet \citep{Sleep1988,Kellogg1999,Zhong2003}.

The thermal structure of a stratified mantle is markedly different to a mantle that is convecting as a whole. The boundary between the two layers restricts the cooling of the lower layer because heat is only allowed to conduct across the interface and this is relatively inefficient compared to advection \citep{McNamara2000}. The large difference in temperature between the two layers is expected to be enhanced by the preferential partitioning of heat-producing elements into the lower layer \citep{Kellogg1999,McNamara2000}, as these elements are generally incompatible \citep{Knittle1998,Blundy2003}. 

\begin{figure}
    \centering
    \includegraphics[width=\hsize]{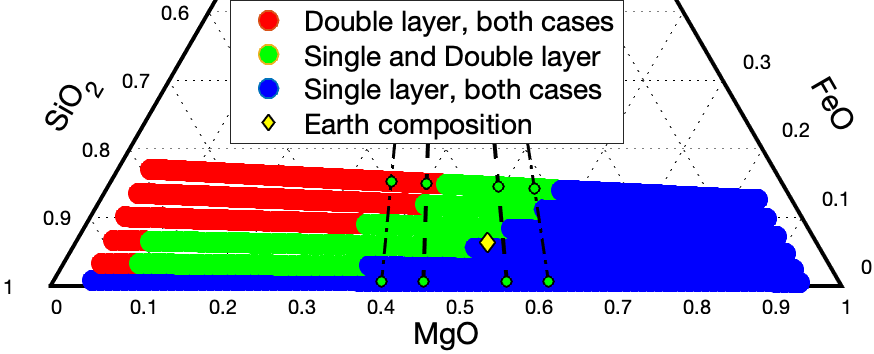}
    \caption{Overview of stable stratification of the mantle as a function of composition for a case with $K_D=0.35$ and $\Delta \rho_{crit}=225$ kg m$^{-3}$ and a case stimulating layering, with $K_D=0.1$ and $\Delta \rho_{crit} = 150$ kg m$^{-3}$. Red dots indicate stable stratification in both cases, blue dots in neither case, and green dots only in the latter case. Composition is shown in terms of molar FeO, MgO and SiO$_2$ abundances, for a range of Mg/Si from 0.05 to 20, and FeO from 1 wt\% to 20 wt\%. The 1$\sigma$ and 2$\sigma$ ranges of Mg/Si-ratios from the Hypatia catalog are shown as vertical dashed and dot-dashed lines, respectively. Green dots on these lines correspond to compositions shown in Table \ref{tab:comps}. The composition of the Earth mantle from \cite{McDonough2003} is shown as a yellow diamond.}
    \label{fig:Triangle_comparison}
\end{figure}

The range of bulk mantle compositions that can lead to stable stratification depends on the style of crystallization, the iron partitioning coefficient $K_D$, and the critical density contrast for stable stratification $\Delta\rho_{crit}$. If we relax our assumption of homogeneity in the upper mantle cumulates, then a smaller volume of more iron-enriched cumulates would form, which would lead to mantle stratification in virtually all compositional cases. This would also be the case for the Earth's composition, which is inconsistent with seismic evidence showing that we have no such global and thick FeO-rich layer \citep{Helffrich2001}. This can either be explained by melting of the FeO-rich cumulates as they sink through the mantle \citep{Ballmer2017_MO} or by a lesser degree of iron enrichment due to chemical re-equilibration of upper mantle cumulates, as we assume here.

In our models, we consider a conservative estimate for $\Delta\rho_{crit}$, and a lower bound for $K_D$. Other authors find stable stratification for values of $\Delta\rho_{crit}$ as low as 100-150 kg/m$^3$ \citep{Tosi2013,Li2014}. Meanwhile, considering the effect of pressure on $K_D$ would increase its value to 0.3--0.4 at shallow lower mantle pressures (30-50 GPa; \cite{Nakajima2012,Muir2016}), thereby decreasing iron enrichment in the cumulates. Figure \ref{fig:Triangle_comparison} shows the range of mantle stratification behavior when considering end-members in terms of $K_D$ and $\Delta\rho_{crit}$. One end-member considers a low $K_D$ and low $\Delta\rho_{crit}$, promoting iron enrichment in the liquid and related layering, while the other end-member considers a high $K_D$ and high $\Delta\rho_{crit}$, reducing the iron enrichment of the liquid. Exploring these end-members reveals a wide range of exoplanet compositions, where the propensity to stratification is uncertain. However, in the limit of our assumptions, even the conservative end-member predicts mantle layering for a non-negligible subset of exoplanets within the compositional range from the Hypatia catalogue. Additionally, considering Si partitioning into the core would increase mantle Mg/Si, which would decrease the range of bulk planet compositions that lead to a stably stratified mantle. However, the cases which lead to stable stratification are expected to have smaller core sizes, limiting this effect (see Sect.\ \ref{ssec:Metho_composition}).

\subsection{Dynamic regime} \label{sec:Disc_Regime}
The dynamic regime affects the long-term evolution of the coupled interior--atmosphere system. The average mantle temperatures for stagnant lid planets are always higher than for tectonically active planets at a given age, regardless of composition or the presence of stable stratification. The thickness and stability of the lid critically reduces the cooling efficiency of the planet, leading to a hotter interior \citep[e.g.,][]{Reese1998}. However, in stagnant-lid planets, (latent) heat may also efficiently be removed through melting and related extrusive volcanism \citep{OReilly1981,Keller2009,Ogawa2011,Nakagawa2012,Armann2012,Moore2013,Foley2018}. This would cool the interior more efficiently than we have modeled here, reducing the temperature difference between stagnant-lid planets and tectonically active planets. However, average mantle temperatures are still expected to be several hundreds of Kelvin higher for SL than for PT planets \citep{Foley2018}.

The dynamic regime also has a first-order effect on the evolution of volatile reservoirs, since it controls the volatile fluxes (i.e., outgassing and ingassing of the mantle). Planets in the stagnant lid regime are assumed to have a high intrusion-to-extrusion ratio \citep{Crisp1984}, which delays atmospheric build-up compared to plate tectonics. Recycling of lithospheric material into the mantle by delamination allows for significant transport of volatiles from this intruded reservoir in the lithosphere back to the mantle. However, the lack of a transport mechanism from the atmosphere to the interior allows the atmospheric reservoir to grow until melting in the interior, and therefore outgassing, ceases. Thus, all SL-cases reach a steady state where most volatiles are stored in the atmosphere, with a final interior volatile budget depending on the mantle Mg/Si-ratio. In contrast, both ingassing and outgassing are efficient for planets in the plate tectonics regime, and a quasi-equilibrium is reached over timescales of 1--2 Gyr, and controlled by mantle temperature. In turn, mantle thermal evolution is controlled by interior viscosity, which mostly depends on bulk Mg/Si.

While we explore all parameter combinations for both dynamic regimes, the regime itself may actually depend on planetary interior parameters, including bulk composition. Parameters that can influence the propensity of plate tectonics are mantle temperature \citep{vanHeck2011,Noack2014,Weller2015}, rock hydration \citep{Korenaga2010}, planet size \citep{Valencia2009,Korenaga2010,vanHeck2011} and the effects of complex rheologies and depth-dependent parameters \citep{Tackley2013}. Furthermore, the dynamic regime depends on mantle viscosity \citep{Korenaga2010,vanHeck2011_PhD}, which is strongly affected by bulk composition. Plate tectonics may be prevented by an overabundance of FeO or alkalis (Na and K) in the mantle \citep{Stamenkovic2016,Unterborn2017_Tect}, or of a globally thick water layer at the surface, which would suppress decompressional mantle melting and prevent outgassing \citep{Kite2009}. Additionally, core size has a minor effect on dynamic regime, with the likelihood of forming plate tectonics increasing with increasing core size \citep{Noack2014Habit}. It does not decrease significantly with smaller core sizes than Earth, so for our parameter range, core size does not have a significant effect. Lastly, the tectonic regime may also be affected by surface temperature \citep{Lenardic2008,Landuyt2009,Noack2012,Gillmann2014}, which is directly linked to atmospheric size and composition (and thereby perhaps interior composition). Thus, there may be a limited range of bulk planet compositions which can lead to long-lived plate tectonics. These ranges need to be better quantified in the future in order to link any upcoming characterizations of exoplanetary atmospheres with the dominant dynamic regime (PT or SL) and mantle composition (Fig.\ \ref{fig:Res_Vol}). 

\subsection{Volatile budget} \label{ssec:Disc_volatiles}
Terrestrial planets can have a wide range of bulk volatile budgets \citep{Raymond2004}, ranging from virtually volatile-depleted planets to planets that contain enough water to form a thick ice layer on top of a rocky mantle \citep{Unterborn2018}. The volatile budget mainly depends on the distance of the planet from the star, stellar surface temperature, and dynamical history of the planetary system \citep{Raymond2004}. Both bulk volatile budget and the distribution of volatiles between the surface and the interior are poorly constrained for Earth and other terrestrial planets. Magma ocean models predict that 70-90\% of the volatile budget outgasses to form a primary atmosphere, depending on the bulk volatile budget \citep{Elkins2008_water}. Meanwhile, Earth currently has most of its volatiles in the interior, with one Earth ocean mass on the surface, and between one ocean mass \citep[e.g.][]{Houser2016,Matsuno2017} to six ocean masses \citep{Bercovici2003,Ohtani2004,Pearson2014} in the mantle. Dynamic melt trapping has been proposed as a way to retain melt in the mantle during magma ocean crystallization, since it could retain up to 77\% of the bulk water and up to 12\% of the bulk CO$_2$ in the bulk mantle \citep{Hier2017}. Here, we find a more volatile-depleted mantle since our modeled mantle is inefficient at retaining water, which could be because we do not treat the mantle transition zone as a water filter \citep{Bercovici2003}. 

Our models predict robust differences between planets in the PT and SL regimes: planetary mantles in the SL regime essentially lose all their volatiles within 4 Gyr through progressive outgassing to the atmosphere. In contrast, planetary mantles in the PT regime, reach a balance (more or less) between outgassing and ingassing after $\sim$3 Gyr (Fig.\ \ref{fig:Res_Vol}) with a significant interior reservoir of volatiles, although initial outgassing is much more efficient than in the SL regime. The regime leaves a characteristic atmospheric signal that may be observable \citep{Foley2016,Tosi2017} in a planet's early life (up to 3 Gyr) and later life (5 Gyr onwards). Similarly, our results indicate an atmospheric signal that is sensitive to bulk Mg/Si, where secondary atmosphere masses are consistently larger for planets with low Mg/Si. However, our interior and volatile models are not tightly coupled, even though volatile exchange and interior evolution affect each other. Water changes rheological properties of the mantle, which causes feedback loops between volatile exchange and evolution of interior temperature \citep{Crowley2011,Nakagawa2015,Nakagawa2017}. This coupling may enhance the effects of interior composition on atmospheric evolution. Future research is needed to explore any related trade-offs, and better quantify the dependency of atmospheric evolution on bulk composition.

While our treatment of outgassing follows robust methods, treatments of ingassing vary throughout the literature. Ingassing in the SL regime is often ignored \citep[e.g.,][]{Noack2014Habit,Tosi2017} or based on delamination (\citep{Foley2018}, this study). For planets in the PT regime, our models are simplified, such that ingassing rates do not depend on Mg/Si or temperature (cf. \cite{Noack2014Habit}). The net ingassing rate may increase with decreasing mantle temperature because of hydrous phase stability \citep{Frost2006,Iwamori2007}, but this would only enhance the differences between low- and high-Mg/Si planets in our results.

We also ran an extended set of models for a large range of bulk-volatile budgets and we confirm that they predict a similar evolutionary trend as the models presented in Figure 9. Thus, any predicted differences in atmospheric size based on interior properties trade off with bulk planet volatile budget. Another ambiguity is shown in both PT and SL resulting in similar secondary atmosphere reservoir sizes between 3 and 5 Gyr. However, the regimes have distinct and characteristic evolutionary paths, with the PT regime characterized by very strong initial outgassing followed by a quasi-balance, where ingassing dominates over outgassing for most compositions, and the SL regime characterized by slow but persistent outgassing. This distinction in evolutionary pathway may help to distinguish between the regimes as additional volatiles with different outgassing behavior are considered. In this study, we consider water as the sole (and ``representative'') volatile in our modeling, but other studies also consider CO$_2$ since it is a major greenhouse gas \citep[e.g.,][]{Sleep2001,Franck2002,Lebrun2013,Tosi2017,Dorn2018_outgassing,Foley2018}. Including the whole range of carbon volatile species (such as CO$_2$ and methane) in coupled interior--atmosphere models should provide additional testable constraints since their behavior is strongly dependent on the oxygen fugacity of the mantle \citep{Mysen2011}. Furthermore, the presence of carbon volatile species can suppress the mantle solidus, but in a very distinct way, as compared to water \citep{Wyllie1976,Dasgupta2010}. Their outgassing and ingassing behavior is also different from that of water \citep{Dasgupta2013,Tosi2017}, which likely results in different evolutionary tracks for water and carbon species in the atmosphere, providing complementary information about the interior.  Other volatile species may also serve as important outgassing products that carry complementary information about the interior, such as those of sulfur and nitrogen \citep{Gaillard2014,Laneuville2018}. 

\section{Conclusions} \label{sec:Conc}
The atmospheric characterization of terrestrial exoplanets has the potential to provide indirect constraints on the interior of the planet as a function of its bulk composition. Bulk planet composition is controlled by stellar composition and it affects atmospheric evolution by changing the balance between interior outgassing and ingassing over the course of long-term evolution. We have developed a framework for understanding the key controls of bulk composition on the coupled interior-atmosphere evolution from the magma ocean stage through to the long-term cycling of volatiles for a planet of Earth-like mass, radius, and core size. We explore the effects of varying exoplanet compositions in terms of major rock forming compounds (i.e., MgO-FeO-SiO$_2$ system) on the structure and evolution of the planetary interior and the exchange of water between the interior and the atmosphere. Our results demonstrate that the bulk planetary composition affects interior evolution and produces a diversity of atmosphere sizes. Future spectroscopic analyses of exoplanetary atmospheres may be able to constrain bulk planet composition, especially when considering more volatile species. The main systematic effects of bulk composition on coupled interior-atmosphere evolution, as found with our model, are:
\begin{itemize}
    \item Depending on the host star abundance, we find that around half of terrestrial exoplanets have compositionally stratified mantles with double-layered convection, unlike Earth. Increasing mantle FeO content and decreasing Mg/Si increases the propensity of forming stably stratified mantles. 
    
    \item Stable mantle stratification reduces the cooling rate of the interior and thereby increases lower-mantle temperature, but does not influence the size of the atmospheric reservoir as strongly as the difference between the plate tectonics and stagnant lid regimes.
    
    \item Planets in the stagnant lid regime tend to have hotter interiors and more massive atmospheres later in the planet's life. Meanwhile, planets in the plate tectonics regime have more efficient outgassing, and therefore form more massive atmospheres early in a planet's life. The difference between the plate tectonics and stagnant lid regimes has a first-order effect on atmospheric mass (and likely on volatile speciation) for the majority of a planet's lifetime, and is potentially observable.
    
    \item The bulk Mg/Si-ratio is the main compositional control on mantle viscosity, where viscosity increases with decreasing Mg/Si. Therefore, planets with high bulk Mg/Si-ratios cool faster than planets with low Mg/Si-ratios. 
    
    \item Directly related to its effects on mantle viscosity, bulk Mg/Si has a second-order effect on mantle outgassing (with low Mg/Si promoting outgassing). Mg/Si thus controls the competition between outgassing and ingassing (and hence atmospheric mass) for planets in the plate tectonics regime. In terms of planets in the stagnant lid regime, Mg/Si controls the fraction of volatiles that outgasses to the atmosphere, and that remains in the mantle. The lid reservoir is small regardless of composition.
    
\end{itemize}

\begin{acknowledgements}
RS has been funded by ETH grant number ETH-18 18-2. DJB acknowledges Swiss National Science Foundation (SNSF) Ambizione Grant 173992. CD acknowledges support from the Swiss National Science Foundation under grant PZ00P2\_174028. This work was in part carried out within the frame of the National Center for Competence in Research {\it PlanetS}. The authors would like to thank the editor and an anonymous reviewer for their helpful feedback and comments that improved the manuscript. We also thank S.\ Mojzsis and C.\ Unterborn for their useful feedback during discussions. 
\end{acknowledgements}

\bibliographystyle{aa}
\bibliography{Thesis_paper}

\begin{thebibliography}{159}
\expandafter\ifx\csname natexlab\endcsname\relax\def\natexlab#1{#1}\fi

\bibitem[{Abe(1997)}]{Abe1997}
Abe, Y. 1997, {Phys.\ Earth Planet.\ Inter.}, 100, 27

\bibitem[{Andrault {et~al.}(2011)Andrault, Bolfan-Casanova, \&
  Nigro}]{Andrault2011}
Andrault, D., Bolfan-Casanova, N., \& Nigro, G. e.~a. 2011, Earth Planet.\
  Sci.\ Lett., 304, 251

\bibitem[{Andrault {et~al.}(2018)Andrault, Pesce, Manthilake, Monteux,
  Bolfan-Casanova, Chantel, Novella, Guignot, King, Iti{\'e},
  {et~al.}}]{Andrault2018}
Andrault, D., Pesce, G., Manthilake, G., {et~al.} 2018, Nat.\ Geosci.\, 11, 139

\bibitem[{Andrault {et~al.}(2012)Andrault, Petitgirard, \&
  Nigro}]{Andrault2012}
Andrault, D., Petitgirard, S., \& Nigro, G. e.~a. 2012, Nat., 497, 354–357

\bibitem[{Armann \& Tackley(2012)}]{Armann2012}
Armann, M. \& Tackley, P.~J. 2012, J.\ Geophys.\ Res.: Planets, 117

\bibitem[{Asplund {et~al.}(2005)Asplund, Grevesse, \& Sauval}]{Asplund2005}
Asplund, M., Grevesse, N., \& Sauval, A. 2005, in ASPCS, Vol. 336, {Cosmic
  Abundances as Records of Stellar Evolution and Nucleosynthesis}, ed.
  T.~Barnes, III \& F.~Bash, 25pp.

\bibitem[{Asplund {et~al.}(2009)Asplund, Grevesse, Sauval, \&
  Scott}]{Asplund2009}
Asplund, M., Grevesse, N., Sauval, A.~J., \& Scott, P. 2009, ARA\&A, 47, 481

\bibitem[{Aubaud {et~al.}(2004)Aubaud, Hauri, \& Hirschmann}]{Aubaud2004}
Aubaud, C., Hauri, E.~H., \& Hirschmann, M.~M. 2004, Geophys.\ Res.\ Lett., 31

\bibitem[{Ballmer {et~al.}(2017{\natexlab{a}})Ballmer, Houser, \&
  Hernlund}]{Ballmer2017_silica}
Ballmer, M.~D., Houser, C., \& Hernlund, J. e.~a. 2017{\natexlab{a}}, Nat.\
  Geosci., 10, 236

\bibitem[{Ballmer {et~al.}(2017{\natexlab{b}})Ballmer, Louren\c{c}o, \&
  Hirose}]{Ballmer2017_MO}
Ballmer, M.~D., Louren\c{c}o, D.~L., \& Hirose, K. e.~a. 2017{\natexlab{b}},
  Geochem.\ Geophys.\ Geosyst., 18, 22 pp

\bibitem[{Bedell {et~al.}(2018)Bedell, Bean, Meléndez, Spina, Ramírez,
  Asplund, Alves-Brito, Santos, Dreizler, Yong, \& et~al.}]{Bedell2018}
Bedell, M., Bean, J.~L., Meléndez, J., {et~al.} 2018, ApJ, 865, 68

\bibitem[{Bensby {et~al.}(2005)Bensby, Feltzing, Lundstr\"om, \&
  Ilyin}]{Bensby2005}
Bensby, T., Feltzing, S., Lundstr\"om, I., \& Ilyin, I. 2005, A\&A, 433, 185

\bibitem[{Bercovici \& Karato(2003)}]{Bercovici2003}
Bercovici, D. \& Karato, S.-I. 2003, Nat., 425, 39

\bibitem[{Blundy \& Wood(2003)}]{Blundy2003}
Blundy, J. \& Wood, B. 2003, Earth Planet.\ Sci.\ Lett., 210, 383

\bibitem[{Bond {et~al.}(2010)Bond, O'Brien, \& Lauretta}]{Bond2010}
Bond, J.~C., O'Brien, D.~P., \& Lauretta, D.~S. 2010, ApJ, 715, 1050

\bibitem[{Botelho {et~al.}(2019)Botelho, Milone, Mel{\'e}ndez, Bedell, Spina,
  Asplund, dos Santos, Bean, Ram{\'\i}rez, Yong, {et~al.}}]{Botelho2019}
Botelho, R.~B., Milone, A. d.~C., Mel{\'e}ndez, J., {et~al.} 2019, MNRAS, 482,
  1690

\bibitem[{Boukar{\'e} {et~al.}(2018)Boukar{\'e}, Parmentier, \&
  Parman}]{Boukare2018}
Boukar{\'e}, C.-E., Parmentier, E., \& Parman, S. 2018, Earth Planet.\ Sci.\
  Lett., 491, 216

\bibitem[{Boukar\'e {et~al.}(2015)Boukar\'e, Ricard, \& Fiquet}]{Boukare2015}
Boukar\'e, C.~E., Ricard, Y., \& Fiquet, G. 2015, J.\ Geophys.\ Res.: Solid
  Earth, 120, 6085

\bibitem[{{Bower} {et~al.}(2019){Bower}, {Kitzmann}, {Wolf}, Sanan, Dorn, \&
  Oza}]{Bower2019}
{Bower}, D.~J., {Kitzmann}, D., {Wolf}, A.~S., {et~al.} 2019, A\&A., 631, A108

\bibitem[{Bower {et~al.}(2018)Bower, Sanan, \& Wolf}]{Bower2018}
Bower, D.~J., Sanan, P., \& Wolf, A. 2018, {Phys. Earth Planet. Inter.}, 274,
  49

\bibitem[{Brewer \& Fischer(2016)}]{Brewer2016}
Brewer, J.~M. \& Fischer, D.~A. 2016, ApJ, 831, 20

\bibitem[{Brugger {et~al.}(2017)Brugger, Mousis, Deleuil, \&
  Deschamps}]{Brugger2017}
Brugger, B., Mousis, O., Deleuil, M., \& Deschamps, F. 2017, ApJ, 850, 12pp.

\bibitem[{Brugman {et~al.}(2020)Brugman, Philips, \& Till}]{Brugman2020}
Brugman, K., Philips, M., \& Till, C. 2020, abstract for conference "Exoplanets
  in our backyard"

\bibitem[{Carter-Bond {et~al.}(2012)Carter-Bond, O'Brien, \&
  Raymond}]{Carter2012}
Carter-Bond, J.~C., O'Brien, D.~P., \& Raymond, S.~N. 2012, ApJ, 760, 44

\bibitem[{Corgne {et~al.}(2008)Corgne, Keshav, \& Wood}]{Corgne2008}
Corgne, A., Keshav, S., \& Wood, B. e.~a. 2008, Geochim.\ Cosmochim.\ Acta, 72,
  574

\bibitem[{Crisp(1984)}]{Crisp1984}
Crisp, J. 1984, {J.\ Volcanol.\ Geotherm.\ Res.}, 20, 177

\bibitem[{Crowley {et~al.}(2011)Crowley, G{\'e}rault, \&
  O'Connell}]{Crowley2011}
Crowley, J.~W., G{\'e}rault, M., \& O'Connell, R.~J. 2011, Earth Planet.\ Sci.\
  Lett., 310, 380

\bibitem[{Dasgupta \& Hirschmann(2010)}]{Dasgupta2010}
Dasgupta, R. \& Hirschmann, M. 2010, Earth Planet.\ Sci.\ Lett., 298, 1

\bibitem[{Dasgupta {et~al.}(2013)Dasgupta, Mallik, \& et~al.}]{Dasgupta2013}
Dasgupta, R., Mallik, A., \& et~al., K.~T. 2013, Nat., 493, 211

\bibitem[{de~Koker {et~al.}(2013)de~Koker, Karki, \& Stixrude}]{deKoker2013}
de~Koker, N., Karki, B.~B., \& Stixrude, L. 2013, EPSL, 361, 58

\bibitem[{Dixon \& Stolper(1995)}]{Dixon1995}
Dixon, J.~E. \& Stolper, E.~M. 1995, J.\ Petrol., 36, 1633

\bibitem[{Dorn {et~al.}(2015)Dorn, Khan, \& Heng}]{Dorn2015}
Dorn, C., Khan, A., \& Heng, K. e.~a. 2015, A\&A, 577, 18 pp.

\bibitem[{Dorn {et~al.}(2018{\natexlab{a}})Dorn, Mosegaard, Grimm, \&
  Alibert}]{Dorn2018_TRAPPIST}
Dorn, C., Mosegaard, K., Grimm, S., \& Alibert, Y. 2018{\natexlab{a}}, ApJ,
  865, 24pp.

\bibitem[{Dorn {et~al.}(2018{\natexlab{b}})Dorn, Noack, \&
  Rozel}]{Dorn2018_outgassing}
Dorn, C., Noack, L., \& Rozel, A. 2018{\natexlab{b}}, A\&A, 614, 20 pp.

\bibitem[{Elkins-Tanton(2008)}]{Elkins2008}
Elkins-Tanton, L.~T. 2008, Earth Planet.\ Sci.\ Lett., 271, 181

\bibitem[{Elkins-Tanton \& Seager(2008)}]{Elkins2008_water}
Elkins-Tanton, L.~T. \& Seager, S. 2008, ApJ, 685, 1237

\bibitem[{Ferot \& Bolfan-Casanova(2012)}]{Ferot2012}
Ferot, A. \& Bolfan-Casanova, N. 2012, Earth Planet.\ Sci.\ Lett., 349, 218

\bibitem[{Fiquet {et~al.}(2010)Fiquet, Auzende, \& Siebert}]{Fiquet2010}
Fiquet, G., Auzende, A., \& Siebert, J. e.~a. 2010, Science, 329, 1516

\bibitem[{Fischer {et~al.}(2017)Fischer, Campbell, \& Ciesla}]{Fischer2017}
Fischer, R.~A., Campbell, A.~J., \& Ciesla, F.~J. 2017, Earth Planet.\ Sci.\
  Lett., 458, 252

\bibitem[{Foley \& Driscoll(2016)}]{Foley2016}
Foley, B.~J. \& Driscoll, P.~E. 2016, Geochem.\ Geophys.\ Geosyst., 17, 1885

\bibitem[{{Foley} \& {Smye}(2018)}]{Foley2018}
{Foley}, B.~J. \& {Smye}, A.~J. 2018, Astrobiology, 18, 873

\bibitem[{Forte \& Mitrovica(1996)}]{Forte1996}
Forte, A.~M. \& Mitrovica, J.~X. 1996, Geophys.\ Res.\ Lett., 23, 1147

\bibitem[{Franck {et~al.}(2002)Franck, Kossacki, Von~Bloth, \&
  Bounama}]{Franck2002}
Franck, S., Kossacki, K.~J., Von~Bloth, W., \& Bounama, C. 2002, Tellus B, 54,
  325

\bibitem[{Frank {et~al.}(2014)Frank, Meyer, \& Mojzsis}]{Frank2014}
Frank, E.~A., Meyer, B.~S., \& Mojzsis, S.~J. 2014, Icarus, 243, 274

\bibitem[{Frost(2006)}]{Frost2006}
Frost, D.~J. 2006, Rev.\ Mineral.\ Geochem., 62, 243

\bibitem[{Fuentes {et~al.}(2019)Fuentes, Crowley, Dasgupta, \&
  Mitrovica}]{Fuentes2019}
Fuentes, J.~J., Crowley, J.~W., Dasgupta, R., \& Mitrovica, J.~X. 2019, Earth
  Planet.\ Sci.\ Lett., 511, 154

\bibitem[{Fulton {et~al.}(2017)Fulton, Petigura, Howard, Isaacson, Marcy,
  Cargile, Hebb, Weiss, Johnson, Morton, {et~al.}}]{Fulton2017}
Fulton, B.~J., Petigura, E.~A., Howard, A.~W., {et~al.} 2017, AJ, 154, 109

\bibitem[{Gaillard \& Scaillet(2014)}]{Gaillard2014}
Gaillard, F. \& Scaillet, B. 2014, Earth Planet.\ Sci.\ Lett., 403, 307

\bibitem[{Gillmann \& Tackley(2014)}]{Gillmann2014}
Gillmann, C. \& Tackley, P. 2014, J.\ Geophys.\ Res.: Planets, 119, 1189

\bibitem[{Hayashi {et~al.}(1979)Hayashi, Nakazawa, \& Mizuno}]{Hayashi1979}
Hayashi, C., Nakazawa, K., \& Mizuno, H. 1979, Earth Planet.\ Sci.\ Lett., 43,
  22

\bibitem[{Helffrich \& Wood(2001)}]{Helffrich2001}
Helffrich, G.~R. \& Wood, B.~J. 2001, Nat., 412, 501

\bibitem[{Hess \& Parmentier(2001)}]{Hess2001}
Hess, P. \& Parmentier, E. 2001, in Lunar Planet. Sci. Conf. XXXII, Abstr, Vol.
  1319

\bibitem[{Hier-Majumder \& Hirschmann(2017)}]{Hier2017}
Hier-Majumder, S. \& Hirschmann, M.~M. 2017, Geochem., Geophys., Geosyst., 18,
  3078

\bibitem[{Hinkel \& Unterborn(2018)}]{Hinkel2018}
Hinkel, N. \& Unterborn, C. 2018, ApJ, 853, 14pp.

\bibitem[{Hinkel {et~al.}(2014)Hinkel, Timmes, \& Young}]{Hypatia}
Hinkel, N.~R., Timmes, F., \& Young, P. A. e.~a. 2014, AJ, 148, 54

\bibitem[{Hirschmann(2000)}]{Hirschmann2000}
Hirschmann, M.~M. 2000, Geochem.\ Geophys.\ Geosyst., 1

\bibitem[{Holloway \& Blank(1994)}]{Holloway1994}
Holloway, J.~R. \& Blank, J.~G. 1994, Rev.\ Mineral.\ Geochem., 30, 187

\bibitem[{Houser(2016)}]{Houser2016}
Houser, C. 2016, Earth Planet.\ Sci.\ Lett., 448, 94

\bibitem[{Iwamori(2007)}]{Iwamori2007}
Iwamori, H. 2007, Chem.\ Geol., 239, 182

\bibitem[{Javoy {et~al.}(2010)Javoy, Kaminski, \& Guyot}]{Javoy2010}
Javoy, M., Kaminski, E., \& Guyot, F. e.~a. 2010, {Earth Planet.\ Sci.\ Lett.},
  293, 259

\bibitem[{Kamata(2018)}]{Kamata2018}
Kamata, S. 2018, {J.\ Geophys.\ Res.: Planets}, 123, 93

\bibitem[{Katz {et~al.}(2003)Katz, Spiegelman, \& Langmuir}]{Katz2003}
Katz, R., Spiegelman, M., \& Langmuir, C. 2003, Geochem.\ Geophys.\ Geosyst.,
  4, 19 pp.

\bibitem[{Keller \& Tackley(2009)}]{Keller2009}
Keller, T. \& Tackley, P.~J. 2009, Icarus, 202, 429

\bibitem[{Kellogg {et~al.}(1999)Kellogg, Hager, \& Van Der~Hilst}]{Kellogg1999}
Kellogg, L.~H., Hager, B.~H., \& Van Der~Hilst, R.~D. 1999, Science, 283, 1881

\bibitem[{Kesson {et~al.}(1998)Kesson, Gerald, \& Shelley}]{Kesson1998}
Kesson, S., Gerald, J.~F., \& Shelley, J. 1998, Nat., 393, 252

\bibitem[{Kiefer {et~al.}(2015)Kiefer, Filiberto, Sandu, \& Li}]{Kiefer2015}
Kiefer, W.~S., Filiberto, J., Sandu, C., \& Li, Q. 2015, Geochim.\ Cosmochim.\
  Acta, 162, 247

\bibitem[{Kite {et~al.}(2009)Kite, Manga, \& Gaidos}]{Kite2009}
Kite, E.~S., Manga, M., \& Gaidos, E. 2009, ApJ, 700, 1732

\bibitem[{Knittle(1998)}]{Knittle1998}
Knittle, E. 1998, The Core-Mantle Boundary Region, Geodyn. Ser, 28, 119

\bibitem[{Korenaga(2010)}]{Korenaga2010}
Korenaga, J. 2010, ApJ, 725, L43

\bibitem[{Korenaga(2013)}]{Korenaga2013}
Korenaga, J. 2013, Annu.\ Rev.\ Earth Planet.\ Sci., 41, 117

\bibitem[{Labrosse {et~al.}(2007)Labrosse, Hernlund, \& Coltice}]{Labrosse2007}
Labrosse, S., Hernlund, J., \& Coltice, N. 2007, Nat., 450, 866

\bibitem[{Landuyt \& Bercovici(2009)}]{Landuyt2009}
Landuyt, W. \& Bercovici, D. 2009, Earth Planet.\ Sci.\ Lett., 277, 29

\bibitem[{Laneuville {et~al.}(2018)Laneuville, Kameya, \&
  Cleaves}]{Laneuville2018}
Laneuville, M., Kameya, M., \& Cleaves, H. 2018, Astrobiology, 18, 897

\bibitem[{Lebrun {et~al.}(2013)Lebrun, Massol, \& Chassefi\`ere}]{Lebrun2013}
Lebrun, T., Massol, H., \& Chassefi\`ere, E. e.~a. 2013, {J.\ Geophys.\ Res.:
  Planets}, 118, 1155

\bibitem[{L{\'e}ger {et~al.}(2011)L{\'e}ger, Grasset, Fegley, Codron, Albarede,
  Barge, Barnes, Cance, Carpy, Catalano, {et~al.}}]{Leger2011}
L{\'e}ger, A., Grasset, O., Fegley, B., {et~al.} 2011, Icarus, 213, 1

\bibitem[{Lenardic {et~al.}(2008)Lenardic, Jellinek, \& Moresi}]{Lenardic2008}
Lenardic, A., Jellinek, A., \& Moresi, L.-N. 2008, Earth Planet.\ Sci.\ Lett.,
  271, 34

\bibitem[{Li {et~al.}(2014)Li, Deschamps, \& Tackley}]{Li2014}
Li, Y., Deschamps, F., \& Tackley, P.~J. 2014, Geophys.\ J.\ Int., 199, 914

\bibitem[{Lodders {et~al.}(2009)Lodders, Palme, \& Gail}]{Lodders2009}
Lodders, K., Palme, H., \& Gail, H. 2009, JE Tr{\"u}mper, 4, 44

\bibitem[{Louren{\c{c}}o {et~al.}(2016)Louren{\c{c}}o, Rozel, \&
  Tackley}]{Lourencco2016}
Louren{\c{c}}o, D.~L., Rozel, A., \& Tackley, P.~J. 2016, Earth Planet.\ Sci.\
  Lett., 439, 18

\bibitem[{Matsuno {et~al.}(2017)Matsuno, Suetsugu, Baba, Tada, Shimizu,
  Shiobara, Isse, Sugioka, Ito, Obayashi, {et~al.}}]{Matsuno2017}
Matsuno, T., Suetsugu, D., Baba, K., {et~al.} 2017, Earth Planet.\ Sci.\ Lett.,
  462, 189

\bibitem[{Maurice {et~al.}(2017)Maurice, Tosi, \& Samuel}]{Maurice2017}
Maurice, M., Tosi, N., \& Samuel, H. e.~a. 2017, J.\ Geophys.\ Res.: Planets,
  122, 577

\bibitem[{McDonough(2003)}]{McDonough2003}
McDonough, W. 2003, Treatise on geochemistry, 547

\bibitem[{McDonough \& Sun(1995)}]{McDonough1995}
McDonough, W.~F. \& Sun, S.-S. 1995, Chemical geology, 120, 223

\bibitem[{McNamara \& van Keken(2000)}]{McNamara2000}
McNamara, A. \& van Keken, P. 2000, Geochem.\ Geophys.\ Geosyst., 1, 17pp.

\bibitem[{Mitrovica \& Forte(2004)}]{Mitrovica2004}
Mitrovica, J. \& Forte, A. 2004, Earth Planet.\ Sci.\ Lett., 225, 177

\bibitem[{Moore \& Webb(2013)}]{Moore2013}
Moore, W. \& Webb, A. 2013, Nat., 501, 501

\bibitem[{Moriarty {et~al.}(2014)Moriarty, Madhusudhan, \&
  Fischer}]{Moriarty2014}
Moriarty, J., Madhusudhan, N., \& Fischer, D. 2014, ApJ, 787, 81

\bibitem[{Muir \& Brodholt(2016)}]{Muir2016}
Muir, J.~M. \& Brodholt, J.~P. 2016, Phys. Earth Planet. Inter., 257, 12

\bibitem[{Mysen {et~al.}(2011)Mysen, Kumamoto, Cody, \& Fogel}]{Mysen2011}
Mysen, B., Kumamoto, K., Cody, G., \& Fogel, M. 2011, Geochim.\ Cosmochim.\
  Acta, 75, 6183

\bibitem[{Nakagawa \& Iwamori(2017)}]{Nakagawa2017}
Nakagawa, T. \& Iwamori, H. 2017, J.\ Geophys.\ Res.: Solid Earth, 122, 8431

\bibitem[{Nakagawa {et~al.}(2015)Nakagawa, Nakakuki, \& Iwamori}]{Nakagawa2015}
Nakagawa, T., Nakakuki, T., \& Iwamori, H. 2015, Geochem., Geophys., Geosyst.,
  16, 1449

\bibitem[{Nakagawa \& Tackley(2012)}]{Nakagawa2012}
Nakagawa, T. \& Tackley, P.~J. 2012, Earth Planet.\ Sci.\ Lett., 329, 1

\bibitem[{{Nakajima} \& {Stevenson}(2015)}]{Nakajima2015}
{Nakajima}, M. \& {Stevenson}, D.~J. 2015, Earth Planet.\ Sci.\ Lett., 427, 286

\bibitem[{Nakajima {et~al.}(2012)Nakajima, Frost, \& Rubie}]{Nakajima2012}
Nakajima, Y., Frost, D.~J., \& Rubie, D.~C. 2012, J.\ Geophys.\ Res.: Solid
  Earth, 117

\bibitem[{Noack \& Breuer(2014)}]{Noack2014}
Noack, L. \& Breuer, D. 2014, Planet.\ Space Sci., 98, 41

\bibitem[{Noack {et~al.}(2012)Noack, Breuer, \& Spohn}]{Noack2012}
Noack, L., Breuer, D., \& Spohn, T. 2012, Icarus, 217, 484

\bibitem[{Noack {et~al.}(2014)Noack, Godolt, von Paris, Plesa, Stracke, Breuer,
  \& Rauer}]{Noack2014Habit}
Noack, L., Godolt, M., von Paris, P., {et~al.} 2014, Planet.\ Space Sci., 98,
  14

\bibitem[{Nomura {et~al.}(2011)Nomura, Ozawa, \& Tateno}]{Nomura2011}
Nomura, R., Ozawa, H., \& Tateno, S. e.~a. 2011, Nat., 473, 199

\bibitem[{Ogawa \& Yanagisawa(2011)}]{Ogawa2011}
Ogawa, M. \& Yanagisawa, T. 2011, J.\ Geophys.\ Res.: Planets, 116

\bibitem[{Ohtani {et~al.}(2004)Ohtani, Litasov, Hosoya, \& et~al.}]{Ohtani2004}
Ohtani, E., Litasov, K., Hosoya, T., \& et~al. 2004, Phys. Earth Planet.
  Inter., 143, 255

\bibitem[{O'Reilly \& Davies(1981)}]{OReilly1981}
O'Reilly, T.~C. \& Davies, G.~F. 1981, Geophys.\ Res.\ Lett., 8, 313

\bibitem[{Owen \& Wu(2013)}]{Owen2013}
Owen, J.~E. \& Wu, Y. 2013, ApJ, 775, 105

\bibitem[{Paulson {et~al.}(2005)Paulson, Zhong, \& Wahr}]{Paulson2005}
Paulson, A., Zhong, S., \& Wahr, J. 2005, Geophys.\ J.\ Int., 163, 357

\bibitem[{Pearson {et~al.}(2014)Pearson, Brenker, \& Nestola}]{Pearson2014}
Pearson, D., Brenker, F., \& Nestola, F. e.~a. 2014, Nat., 507, 221

\bibitem[{Plesa {et~al.}(2014)Plesa, Tosi, \& Breuer}]{Plesa2014}
Plesa, A.-C., Tosi, N., \& Breuer, D. 2014, Earth Planet.\ Sci.\ Lett., 403,
  225

\bibitem[{Raymond {et~al.}(2004)Raymond, Quinn, \& Lunine}]{Raymond2004}
Raymond, S.~N., Quinn, T., \& Lunine, J.~I. 2004, Icarus, 168, 1

\bibitem[{Reese {et~al.}(1998)Reese, Solomatov, \& Moresi}]{Reese1998}
Reese, C., Solomatov, V., \& Moresi, L.-N. 1998, J.\ Geophys.\ Res.: Planets,
  103, 13643

\bibitem[{Reese {et~al.}(1999)Reese, Solomatov, \& Moresi}]{Reese1999}
Reese, C., Solomatov, V., \& Moresi, L.-N. 1999, Icarus, 139, 67

\bibitem[{Rogers \& Seager(2010)}]{Rogers2010}
Rogers, L. \& Seager, S. 2010, ApJ, 712, 974

\bibitem[{Rolf \& Tackley(2011)}]{Rolf2011}
Rolf, T. \& Tackley, P. 2011, Geophys.\ Res.\ Lett., 38

\bibitem[{Rubie {et~al.}(2015)Rubie, Jacobson, \& Morbidelli}]{Rubie2015}
Rubie, D.~C., Jacobson, S.~A., \& Morbidelli, A. e.~a. 2015, Icarus, 248, 89

\bibitem[{Ruedas(2017)}]{Ruedas2017}
Ruedas, T. 2017, Geochem.\ Geophys.\ Geosyst., 18, 3530

\bibitem[{R{\"u}pke {et~al.}(2004)R{\"u}pke, Morgan, Hort, \&
  Connolly}]{Rupke2004}
R{\"u}pke, L.~H., Morgan, J.~P., Hort, M., \& Connolly, J.~A. 2004, Earth
  Planet.\ Sci.\ Lett., 223, 17

\bibitem[{Santos {et~al.}(2017)Santos, Adibekyan, Dorn, Mordasini, Noack,
  Barros, Delgado-Mena, Demangeon, Faria, Israelian, {et~al.}}]{Santos2017}
Santos, N., Adibekyan, V., Dorn, C., {et~al.} 2017, A \& A, 608, A94

\bibitem[{Schilling(1991)}]{Schilling1991}
Schilling, J.-G. 1991, Nat., 352, 397

\bibitem[{Seager {et~al.}(2007)Seager, Kuchner, Hier-Majumder, \&
  Militzer}]{Seager2007}
Seager, S., Kuchner, M., Hier-Majumder, C., \& Militzer, B. 2007, ApJ, 669,
  1279

\bibitem[{Sleep(1988)}]{Sleep1988}
Sleep, N.~H. 1988, Geophys.\ J.\ Int., 95, 437

\bibitem[{Sleep(2000)}]{Sleep2000}
Sleep, N.~H. 2000, J.\ Geophys.\ Res.: Planets, 105, 17563

\bibitem[{Sleep \& Zahnle(2001)}]{Sleep2001}
Sleep, N.~H. \& Zahnle, K. 2001, J.\ Geophys.\ Res.: Planets, 106, 1373

\bibitem[{Solomatov(2000)}]{Solomatov2000}
Solomatov, V. 2000, in {Origin of the Earth and Moon}, ed. R.~Canup \&
  K.~Righter (Tucson, AZ: {University of Arizona Press}), 323–338

\bibitem[{Solomatov(2015)}]{Solomatov2015}
Solomatov, V. 2015, in Treatise on Geophysics, 2nd edn., Vol.~9 (Elsevier),
  81--104

\bibitem[{Solomatov \& Stevenson(1993)}]{Solomatov1993}
Solomatov, V.~S. \& Stevenson, D.~J. 1993, J.\ Geophys.\ Res.: Planets, 98,
  5375

\bibitem[{Stacey \& Davis(2008)}]{Stacey2008}
Stacey, F.~D. \& Davis, P.~M. 2008, Physics of the Earth (Cambridge University
  Press)

\bibitem[{Stamenkovi\'c \& Seager(2016)}]{Stamenkovic2016}
Stamenkovi\'c, V. \& Seager, S. 2016, {ApJ}, 825, {18 pp.}

\bibitem[{Stixrude {et~al.}(2009)Stixrude, de~Koker, \& Sun}]{Stixrude2009}
Stixrude, L., de~Koker, N., \& Sun, N. e.~a. 2009, Earth Planet.\ Sci.\ Lett.,
  278, 226

\bibitem[{Tackley {et~al.}(2013)Tackley, Ammann, \& Brodholt}]{Tackley2013}
Tackley, P.~J., Ammann, M., \& Brodholt, J. P. e.~a. 2013, Icarus, 225, 50

\bibitem[{Takeda(1998)}]{Takeda1998}
Takeda, Y.-T. 1998, J.\ Struct.\ Geol., 20, 1569

\bibitem[{Tange {et~al.}(2009)Tange, Takahashi, \& Nishihara}]{Tange2009}
Tange, Y., Takahashi, E., \& Nishihara, Y. e.~a. 2009, J.\ Geophys.\ Res.:
  Solid Earth, 114

\bibitem[{Tateno {et~al.}(2014)Tateno, Hirose, \& Ohishi}]{Tateno2014}
Tateno, S., Hirose, K., \& Ohishi, Y. 2014, J.\ Geophys.\ Res.: Solid Earth,
  119, 4684

\bibitem[{Thiabaud {et~al.}(2015)Thiabaud, Marboeuf, Alibert, Leya, \&
  Mezger}]{Thiabaud2015}
Thiabaud, A., Marboeuf, U., Alibert, Y., Leya, I., \& Mezger, K. 2015, A\&A,
  580, A30

\bibitem[{Tosi {et~al.}(2017)Tosi, Godolt, \& Stracke}]{Tosi2017}
Tosi, N., Godolt, M., \& Stracke, B. e.~a. 2017, A\&A, 605, 21 pp

\bibitem[{Tosi {et~al.}(2013)Tosi, Plesa, \& Breuer}]{Tosi2013}
Tosi, N., Plesa, A., \& Breuer, D. 2013, {J.\ Geophys.\ Res.: Planets}, 118,
  1512

\bibitem[{Unterborn \& Panero(2017)}]{Unterborn2017_MgSi}
Unterborn, C. \& Panero, W. 2017, ApJ, 845, {9 pp.}

\bibitem[{Unterborn {et~al.}(2018)Unterborn, Desch, Hinkel, \&
  Lorenzo}]{Unterborn2018}
Unterborn, C.~T., Desch, S.~J., Hinkel, N.~R., \& Lorenzo, A. 2018, Nat.\
  Astron., 2, 297

\bibitem[{Unterborn {et~al.}(2016)Unterborn, Dismukes, \&
  Panero}]{Unterborn2016}
Unterborn, C.~T., Dismukes, E.~E., \& Panero, W.~R. 2016, ApJ, 819, 32

\bibitem[{{Unterborn} {et~al.}(2017){Unterborn}, {Hull}, \&
  {Stixrude}}]{Unterborn2017_Tect}
{Unterborn}, C.~T., {Hull}, S.~D., \& {Stixrude}, L. e.~a. 2017, in LPI
  Contributions, Vol. 2042, Habitable Worlds 2017: A System Science Workshop,
  4034

\bibitem[{Unterborn {et~al.}(2015)Unterborn, Johnson, \&
  Panero}]{Unterborn2015}
Unterborn, C.~T., Johnson, J.~A., \& Panero, W.~R. 2015, ApJ, 806, 139

\bibitem[{Unterborn \& Panero(2019)}]{Unterborn2019}
Unterborn, C.~T. \& Panero, W.~R. 2019, J.\ Geophys.\ Res.: Planets, 124, 1704

\bibitem[{Valencia \& O'Connell(2009)}]{Valencia2009}
Valencia, D. \& O'Connell, R.~J. 2009, Earth Planet.\ Sci.\ Lett., 286, 492

\bibitem[{{Valencia} {et~al.}(2006){Valencia}, {O'Connell}, \&
  {Sasselov}}]{Valencia2006}
{Valencia}, D., {O'Connell}, R.~J., \& {Sasselov}, D. 2006, \icarus, 181, 545

\bibitem[{{Valenti} \& {Fischer}(2005)}]{Valenti2005}
{Valenti}, J.~A. \& {Fischer}, D.~A. 2005, ApJS, 159, 141

\bibitem[{van Heck \& Tackley(2011)}]{vanHeck2011}
van Heck, H. \& Tackley, P. 2011, Earth Planet.\ Sci.\ Lett., 310, 252

\bibitem[{van Heck(2011)}]{vanHeck2011_PhD}
van Heck, H.~J. 2011, PhD thesis, ETH Zurich

\bibitem[{van Keken {et~al.}(2011)van Keken, Hacker, Syracuse, \&
  Abers}]{vanKeken2011}
van Keken, P., Hacker, B., Syracuse, E., \& Abers, G. 2011, J.\ Geophys.\ Res.:
  Solid Earth, 116, 15pp.

\bibitem[{Wagner {et~al.}(2019)Wagner, Plesa, \& Rozel}]{Wagner2019}
Wagner, F., Plesa, A.-C., \& Rozel, A. 2019, Geophys.\ J.\ Int., 217, 75

\bibitem[{{Wagner} {et~al.}(2011){Wagner}, {Sohl}, \& {Hussmann}}]{Wagner2011}
{Wagner}, F.~W., {Sohl}, F., \& {Hussmann}, H. e.~a. 2011, \icarus, 214

\bibitem[{Wang {et~al.}(2018)Wang, Lineweaver, \& Ireland}]{Wang2018comp}
Wang, H.~S., Lineweaver, C.~H., \& Ireland, T.~R. 2018, Icarus, 299, 460

\bibitem[{Weiss \& Marcy(2014)}]{Weiss2014}
Weiss, L.~M. \& Marcy, G.~W. 2014, ApJ Lett., 783, L6

\bibitem[{Weller {et~al.}(2015)Weller, Lenardic, \& O'Neill}]{Weller2015}
Weller, M., Lenardic, A., \& O'Neill, C. 2015, Earth Planet.\ Sci.\ Lett., 420,
  85

\bibitem[{Wolery \& Sleep(1976)}]{Wolery1976}
Wolery, T.~J. \& Sleep, N.~H. 1976, J.\ Geol., 84, 249

\bibitem[{Wolf \& Bower(2018)}]{Wolf2018}
Wolf, A.~S. \& Bower, D.~J. 2018, Phys. Earth Planet. Inter., 278, 59

\bibitem[{Wolf {et~al.}(2015)Wolf, Jackson, Dera, \& Prakapenka}]{Wolf2015}
Wolf, A.~S., Jackson, J.~M., Dera, P., \& Prakapenka, V.~B. 2015, J.\ Geophys.\
  Res.: Solid Earth, 120, 7460

\bibitem[{Wordsworth {et~al.}(2018)Wordsworth, Schaefer, \&
  Fischer}]{Wordsworth2018}
Wordsworth, R., Schaefer, L., \& Fischer, R. 2018, AJ, 155, 195

\bibitem[{Wyllie \& Huang(1976)}]{Wyllie1976}
Wyllie, P.~J. \& Huang, W.-L. 1976, Contr.\ Miner.\ Petrol., 54, 79

\bibitem[{Xu {et~al.}(2017)Xu, Yamazaki, \& Sakamoto}]{Xu2017}
Xu, F., Yamazaki, D., \& Sakamoto, N. e.~a. 2017, Earth Planet.\ Sci.\ Lett.,
  459, 332

\bibitem[{Yamazaki \& Karato(2001)}]{Yamazaki2001}
Yamazaki, D. \& Karato, S. 2001, {Am.\ Mineral.}, 86, 385

\bibitem[{Zahnle \& Catling(2017)}]{Zahnle2017}
Zahnle, K. \& Catling, D. 2017, ApJ, 843, 23pp.

\bibitem[{Zhong \& Hager(2003)}]{Zhong2003}
Zhong, S. \& Hager, B.~H. 2003, Geophys.\ J.\ Int., 154, 666

\bibitem[{Zou(1998)}]{Zou1998}
Zou, H. 1998, Geochim.\ Cosmochim.\ Acta, 62, 1937

\end{thebibliography}

\end{document}